\documentclass[12pt]{article}
\usepackage[cp1251]{inputenc}
\usepackage{amsfonts}
\usepackage[english]{babel}
\usepackage{eufrak}
\usepackage{setstack}
\usepackage{cite}
\usepackage{amssymb}
\title{Two models of protoplasm microstructure of the living cell in resting state}
\begin{document}
\author{D.V. Prokhorenko, V.V. Matveev
\footnote{Institute of Cytology, 194064 Russian Academy of Sciences,
Saint Petersburg, vladimir.matveev@gmail.com}}

 \maketitle
\begin{abstract}
\maketitle In order to develop the methods of thermodynamic analysis
for the living cell,  two models of protoplasm microstructure of the
living cell in resting state were suggested. Both models are based
on the assumption that the Ling's cell as a statistical mechanics
system is non-ergodic. In the first,Van der Waals model, the
protein-protein interactions, which form the physical basis for the
cell functioning, are considered as a interactions of key
importance. It is postulated that protein molecules are situated in
points of some space lattice (the Ling model of a cell) they
assemble to aggregates at equilibrium state, corresponding to the
dead protoplasm. In the second model we consider protein
conformation at the resting state and conformation changes while the
cell is passing from the resting state to the equilibrium state
(dead protoplasm). The investigation of the models and comparison of
their characteristics showed that the convenient tool to define the
energy minimum of the system under consideration is a Hamiltonian
describing the superfluid Bose gas on protein configuration space.
Our approach allows us to define the thermodynamic features of the
living (at resting state) and dead protoplasm in a new way: in the
first case the system is characterized by the unfolded state of
proteins, in the second case proteins are folded and aggregated.
Obtained results prove the applicability of our approaches for
thermodynamic characteristics of the Ling model of a cell.
\end{abstract}
\sloppy
\section{Introduction}

R. Feynman wrote: "All things are made of atoms, and that everything
that living things do can be understood in terms of the jiggling and
wiggling of atoms." (Feynman, 1963). To move beyond this assertion,
it is necessary to adopt common principles of organization of atoms
and molecules in living systems. These principles, compatible with
the existing analytical apparatus of thermodynamics and statistical
physics, have been formulated by Gilbert Ling (Ling, 2006). The
living cell model created by him was used as a starting point of his
study.

According to Ling, fundamental properties of the living cell are
explained by the single physical factor --- sorption properties of
its proteins. An unfolded (linear) protein molecule binding water
(multilayer adsorption) and \({\rm K \mit}^+\) (in the presence of \({\rm Na \mit}^+\)) under the
control of ATP represents the smallest part (unit) of a living
protoplasm which still keeps the main physical characteristics of
the living cell. Later Matveev (2005) offered to call the unit as a
physiological atom or physioatom.

The main physical state of the phisioatom, and accordingly, the
cells comprising them, is, according to Ling, a resting state. The
physical nature of this state determines, on the Ling's theory, all
forms of biological activity of the cell and therefore the analysis
of this state is a key issue of the physical theory of the living
cell.

Compatibility of the Ling's resting cell organization principles
with analytical methods of modern theoretical physics was first
shown in our previous work (Prokhorenko and Matveev, 2011). The base
of our approach is the fact that the majority of statistical
mechanics systems (including the most realistic ones) are
non-ergodic that was proved by one of us (Prokhorenko, 2009). The
generalized thermodynamic analysis (generalized thermodynamics) of
Ling's cell we proposed (Prokhorenko and Matveev, 2011) allowed us
to explain (in framework of adopted boundary conditions) a number of
physiological phenomena that occur when cell is in activated
(excited) state: exothermicity of transition to excited state,
change of cell volume, folding of natively unfolded proteins (which
determine, by Ling, a main features of the resting state), efflux of
cell \({\rm K \mit}^+\) and wider --- major redistribution of physiologically
important ions between the cell and its environment.  However, we
determined the sign (direction) only of these processes, there were
no numerical evaluations. In other words, the results were obtained
as inequalities. These are the normal features of thermodynamics
(not of the generalized one only): it allows us to obtain general
relations between thermodynamic variables which are independent on a
nature of intermolecular interactions. It is needed to construct
models with desired properties (including microscopic models) to
obtain specific numerical values of thermodynamic variables and then
investigate them by methods of statistical mechanics.

Indeed, inequalities we have determined (Prokhorenko and Matveev,
2011) were based on the postulate of relative entropy maximum for
the resting cell standing in the state of thermodynamic equilibrium
with environment. However, an equilibrium with environment does not
mean the equilibrium state (the absolute maximum of entropy) of a
system; that's why the relative nature of the entropy maximum is
indicated. We consider the resting cell as a system in a steady
non-equilibrium state described by the generalized Gibbs
distribution (Prokhorenko and Matveev, 2011). In other words, the
case is the maximum  among of all the states described by
generalized Gibbs distributions and constructed using the fixed set
of first integrals in the involution. These inequalities give
evidences of the negative determinacy of the second entropy
derivative matrix with respect to parameters describing a system in
a state corresponding to the relative entropy maximum.

Involvement of statistical mechanics methods brings up an issue of
certain properties of the investigated system. Ling's model of the
living cell gives, in our opinion, an interesting material for such
analysis.  So, after construction of the generalized thermodynamics
of the resting state, set out in (Prokhorenko and Matveev, 2011), it
makes sense to turn to some of the structural characteristics of the
investigated system (as it often happened in the history of
thermodynamics and statistical mechanics). In our case, the problem
arises of constructing various models of protoplasm, and their
investigation by various (mostly approximate) methods of theoretical
physics. In this paper, the authors make the first steps in this
direction.

The first model, we call the Van der Waals model, focuses on the
nature of interactions between protein molecules only. As Ling, we
assume that protein molecules in the resting cell embedded at
specific sites in the lattice of a crystal, and the distances
between proteins are so long that interaction between them can be
neglected.  This assumption makes it possible to use the method for
calculating thermodynamic potentials of ideal systems in order to
determine thermodynamic potentials of the resting protoplasm with a
specified structure. In the case of dead protoplasm (the state
opposite to the living resting state), the protein molecules are
associated due to secondary (non-covalent) bonds in large
equilibrium aggregates. In this case, to calculate thermodynamic
potentials we use the formula, we have obtained in this paper, for
corrections of free energy values in the case of formation of large
aggregates.

As part of the Van der Waals model we have obtained (i) numerical
estimation for heat amount released when erythrocyte die, (ii) the
estimation for number of protein aggregates appeared in dead
protoplasm, and (iii) the estimation for fraction of whole cell
volume occupied by these aggregates. All these estimates are in good
(qualitative) agreement with available experimental data.

In our second model, we also assume that protein molecules in the
resting cell embedded at specific sites in the lattice of a crystal,
but at this time the focus is on internal structure of a protein.
We consider this model at zero temperature (the energy scale), which
makes it possible to use (to calculate the ground state) some
effective Hamiltonian describing a superfluid Bose gas in the
configuration space of a protein molecule. Based on the
representations set forth in (Prokhorenko and Matveev, 2011), we
define parameters of effective Hamiltonian corresponding to living
and dead states of protoplasm; we show (in accordance with our
assumptions) that proteins in the resting state (that determine key
properties of the system) are natively unfolded, whereas in dead
protoplasm the same protein molecules are folded.

In Appendix 1 the process when the living protoplasm transforms into
a dead one is considered as it appears in the physical point of
view. In Appendix 2 we discuss the mechanism by which ATP is able to
effectively influence sorption properties of proteins of Ling's
model for water and physiologically important cations.

\section{Non-Ergodicity and Crystallization}

Let's consider some non-trivial issues that arise when we consider
the cell as a nonergodic system. According to the Ling's model
(Ling, 2001), water in the resting cell is in a bound
quasi-crystalline state (important, water content is about 44
mole/kg wet weight of the cell). The bound state of water and its
massive amount in the cell has fundamental importance for the
understanding of physiological processes (Ling, 1997). Therefore,
one of key issues of cell physics is the question: which properties
the crystal has in terms of our recently proposed approach,
generalized thermodynamics (Prokhorenko and Matveev, 2011) Let's
begin the consideration of this issue with finding out the relation
between non-ergodicity of a system (for example, the Ling's cell)
and its solidifying capability at low temperatures.

Despite this problem definition is non-physiological, its solution
will allow us to verify once more that a large number of systems of
statistical mechanics, including our model, have a non-ergodicity
property. Our argument will be largely heuristic rather than
rigorously mathematical character. In the mathematical physics the
rigorous theory is often preceded by formal theories handling
objects of poorly ascertained mathematical context. However, the
heuristics presented here are of interest, in our view, as a basis
for more rigorous methods.

At first, let's define the ergodicity for statistical mechanics
systems.

\textbf{Definition.} Suppose the quantum system is described by
Hamiltonian \(H\) and \(K_1,...,K_l\) are some commuting (among
themselves) self-adjoint integrals of motion. The system is called
ergodic with respect to the set of integrals \(K_1,...,K_l\) if any
dynamical variable commuting with \(H,\;K_1,...,K_l\) is their
function.

To give a classical analog of this definition we should just replace
the word "commutator" by a Poisson bracket everywhere.

Usually, the operator of system momenta \(\vec{P}\) and operator of
particles number \(N\) are used as trivial integrals.

At first let's show how the non-ergodicity of a system arises from
its solidifying capability at low temperatures. Let's consider the
system at solidifying temperatures, supposing them \(T<T_0\)
\(T_0>0\). In addition, the system can move through the space as a
solid body and its coordinates (as as solid body) are six real
numbers
\begin{eqnarray}
x_1,\;x_2,\;x_3\;\varphi_1,\;\varphi_2,\;\varphi_3,
\end{eqnarray}
where \(x_1,\;x_2,\;x_3\) are Cartesian coordinates of the system's
center of mass, and \(\varphi_1,\;\varphi_2,\;\varphi_3\) are some
coordinates characterizing the position of a system (as a solid
body) relative to its center of mass, for example, Euler angles. Let
\(p_1,..p_3,\pi_1,...,\pi_3\) be momenta canonically conjugated to
them. The Hamiltonian of the whole system
\begin{eqnarray}
\hat{H}(x_1,...,x_3,\varphi_1,...,\varphi_3,p_1,...,p_3,\pi_1,...,\pi_3)
\end{eqnarray}
is a function of variables
\(x_1,\;x_2,\;x_3\;\varphi_1,\;\varphi_2,\;\varphi_3\), conjugated
momenta to them and operator "in other variables". Free energy of a
system is given by:
\begin{eqnarray}
F(x_1,...,x_3,\varphi_1,...,\varphi_3,p_1,...,p_3,\pi_1,...,\pi_3|T)\nonumber\\
=-T {\ln \rm tr \mit}
e^{-\frac{\hat{H}(x_1,...,x_3,\varphi_1,...,\varphi_3,p_1,...,p_3,\pi_1,...,\pi_3)}{T}},
\end{eqnarray}
where trace is taken by Hilbert space which is "left" after
separation of variables describing the motion of a system as a solid
body. We won't refine the meaning of words enclosed in quotation
marks considering them as intuitive clear. It's clear that
\(F(x_1,...,x_3,\varphi_1,...,\varphi_3,p_1,...,p_3,\pi_1,...,\pi_3|T)\)
 does not depend on variable \(\varphi_3\) (if, for example,
 \(\varphi_1,\;\varphi_2,\;\varphi_3\) are Euler angles).

But
\begin{eqnarray}
{\ln \rm tr \mit}
e^{-\frac{\hat{H}(x_1,...,x_3,\varphi_1,...,\varphi_3,p_1,...,p_3,\pi_1,...,\pi_3)}{T}}=\sum
\limits_{i=0}^{\infty}d_i(x_1,...,\pi_3)e^{-\frac{\lambda_i(x_1,..,\pi_3)}{T}},
\end{eqnarray}
where \(d_i={\rm dim \mit} L_i\) is a dimension of eigenspace
\(L_i\) of operator \(\hat{H}(x_1,...,\pi_3)\) corresponding to
\(\lambda_i\) eigenvalue of operator \(\hat{H}(x_1,...,\pi_3)\). But
since \(\sum
\limits_{i=0}^{\infty}d_i(x_1,...,\pi_3)e^{-\lambda_i(x_1,..,\pi_3)\beta}\)
(\(\beta:=\frac{1}{T}\)) does not depend on \(\varphi_3\) (for
different \(\beta\)), then \(d_i(x_1,...,\pi_3)\) and
\(\lambda_i(x_1,..,\pi_3)\) are also independent of \(\varphi_3\).
Indeed \(\sum
\limits_{i=0}^{\infty}d_i(x_1,...,\pi_3)e^{-\lambda_i(x_1,..,\pi_3)\beta}\)
is really a Laplace transform of a measure
\begin{eqnarray}
\sum
\limits_{i=0}^{\infty}\delta(\lambda-\lambda_i(x_1,..,\pi_3))d_i(x_1,...,\pi_3).
\end{eqnarray}
So, for all values \(\varphi_3\) when others values parameters
\(x_1,...,\pi_3\) are the same, operators
\(\hat{H}(x_1,...,x_3,\varphi_1,...,\varphi_3,p_1,...,p_3,\pi_1,...,\pi_3)\)
are unitary equivalent. And after making the appropriate unitary
transformation of \(\mathcal{H}\)  depending on \(x_1,...,\pi_3\),
we can conclude that
\(\hat{H}(x_1,...,x_3,\varphi_1,...,\varphi_3,p_1,...,p_3,\pi_1,...,\pi_3)\)
do not depend on \(\varphi_3\). We have considered variables
\(x_1,...,x_3,\varphi_1,...,\varphi_3,p_1,..p_3,\pi_1,...,\pi_3\) as
classical ones since they describe macroscopic degrees of freedom
and appear to be very large. Now we again consider \(\varphi_3\),
\(\pi_3\) as quantum variables, replacing them by corresponding
operators  \(\hat{\varphi_3}\), \(\hat{\pi_3}\),  and considering
the Hamiltonian
\(\hat{H}_1(x_1,...,x_3,\varphi_1,\varphi_2,p_1,...,p_3,\pi_1,\pi_2)\)
obtained by replacing the variables \(\varphi_3\), \(\pi_3\) with
the corresponding quantum-mechanical operators \(\hat{\varphi_3}\),
\(\hat{\pi_3}\), in \(\hat{H}\) to describe our system. This
operator acts in Hilbert space \(\mathcal{H}\otimes\Gamma\) where
\(\Gamma\) is a Hilbert space corresponding to operators
\(\hat{\varphi_3}\), \(\hat{\pi_3}\). The fact of independence of
\(\hat{H}(x_1,...,x_3,\varphi_1,...,\varphi_3,p_1,...,p_3,\pi_1,...,\pi_3)\)
of \(\varphi_3\) is stated now as commutativity of \(\hat{H}_1\)
with \(\hat{\pi_3}\), and the presence of nontrivial first integral
of a system means of course the Hamiltonian degeneracy. The last
fact can indicate the non-ergodicity of the system, but this new
integral should be commutative with the momenta operator. It can be
achieved for example by consideration of the integral
\(\Pi:=\frac{\hat{\pi_3}}{G}\)  instead of \(\hat{\pi_3}\), where
\(G\) behaves like \(\sim V^{\frac{5}{3}}\) if the volume \(V\) of
the system approaches infinity. Then \(\Pi\) asymptotically
commutates with the momenta operator which means the non-ergodicity
of the system. The reason for choosing \(G\sim V^{\frac{5}{3}}\) is
shown below.

However, instead of considering \(\Pi\) as a new independent
integral we prefer other way. Variables
\(x_1,...,x_3,\varphi_1,...,\varphi_3,p_1,...,p_3,\pi_1,...,\pi_3\)
are canonically conjugated variables satisfying to the Hamilton's
evolution at temperatures \(T<T_0\) for the Hamiltonian
\(F(x_1,...,\pi_3|T)\) (see section 3). Just as we did before, we
can show that \(F(x_1,...,\pi_3|T)\) does not depend on
\(x_1,...,x_3,\varphi_1,...,\varphi_3\). But the system in the
equilibrium state has the energy (in the thermodynamic limit)
proportional to the volume of this system, therefore \(E\leq CV\)
for some constant \(C\). On the other hand, if \(I_1,I_2,I_3\) are
the eigenvalues of the inertia operator of our system as a solid
body and \(\omega_1,\omega_2,\omega_3\) are components of angular
velocity along corresponding principal axes of inertia operator,
then \(E\geq\frac{I_1}{2} \omega_1^2+\frac{I_2}{2}
\omega_2^2+\frac{I_3}{2} \omega_3^2\). However, \(I_1,I_2,I_3\sim
V^{\frac{5}{3}}\). Therefore, in the thermodynamic limit
\(\omega_1=\omega_2=\omega_3=0\) and our system can moves only
parallel to itself in other words,
\(\dot{\varphi}_1=\dot{\varphi}_2=\dot{\varphi}_3=0\). So, variables
\(\varphi_1,\;\varphi_2,\;\varphi_3\) are motion integrals of the
system commutating with the impulse operator which makes our system
non-ergodic one.

Since \(CV\geq E\geq\frac{I_1}{2} \omega_1^2+\frac{I_2}{2}
\omega_2^2+\frac{I_3}{2}\omega_3^2=\frac{\pi_1^2}{2I_1}+...+\frac{\pi_3^2}{2I_3}\),
and \(I_1,I_2,I_3\sim V^{\frac{5}{3}}\) then
\(\pi_1.\;\pi_2,\;\pi_3\sim V^{\frac{4}{3}}\).

At low angular velocities \(\omega_1,...,\omega_3\) the free energy
of the whole system is presented as
\begin{eqnarray}
F(x_1,...,\pi_3)=F_0(x_1,...,p_3)+\frac{I_1\omega_1^2}{2}+...+\frac{I_3\omega_3^2}{2},
\end{eqnarray}
for some function \(F_0(x_1,...,p_3)\) of variables \(x_1,...,p_3\).
Momentum variables \(\pi_1,...,\pi_3\) can be chosen so that time
rates of change of canonically conjugated coordinates may be equal
to \(\omega_1,...,\omega_3\). Then
\(\pi_1=I_1\omega_1\),...,\(\pi_3=I_3\omega_3\) and the effective
Hamiltonian of the system equals to
\begin{eqnarray}
F(x_1,...,\pi_3)=F_0(x_1,...,p_3)+\frac{\pi_1^2}{2I_1}+...+\frac{\pi_3^2}{2I_3}.
\end{eqnarray}

The form of this Hamiltonian together with the fact that in the
thermodynamic limit \(\pi_1.\;\pi_2,\;\pi_3\sim V^{\frac{4}{3}}\)
imply that \(\varphi_1,\;\varphi_2,\;\varphi_3\) are the integrals
of motion commutating with an operator of the total system momenta.

Now let us ask the question arising out of the context of our
analysis: why the crystalline state of matter is stable at
temperatures different from zero. The fact that matter solidifies at
zero temperature is almost evident: configuration of the system must
achieve the minimum potential energy. The question arises: why do
atoms keep on doing just small oscillations around points of the
lattice at non-zero temperature and why does the lattice remain
faultless, though thermal fluctuations seem to break it. This
question is closely related to the question of why Ling's cell is
stable at relatively high temperatures. We shall try to answer with
help of our generalized thermodynamics (Prokhorenko and Matveev,
2011).

So, let \(\mathcal{H}\) be a Hilbert space of our system,
\(\hat{H}\) is a Hamiltonian of our system, and \(E_0\) is the
lowest number belonging to spectrum, and \(\hat{E}_0\) is a spectral
projection of \(\hat{H}\) onto the eigensubspace \(\hat{H}\)
corresponding to the eigenvalue \({E}_0\).

In classical terms, at \(T=0\) atoms composing the system have an
arrangement which meets the condition of minimum potential energy.
This means the body solidifies at \(T=0\). As we suppose, at that
moment atoms are situated in points of a crystal lattice. But
according to the accepted approach this lattices are not invariant
under infinitesimal rotation, i.e. the system state obtained from
the initial one by an infinitesimal rotation does not align with the
initial one. This results degeneracy of \(E_0\), i.e. \({\rm tr
\mit}\hat{E}_0>1\).

Now let's complete \(\{\hat{H}\}\) to obtain the complete set of
(commuting) observed values by self-adjoint operators
\(\hat{K}_1,\hat{K}_2,...\). Here we use Dirac terminology (Dirac,
1958).

Completeness of the system of observables
\(\hat{H},\hat{K}_1,\hat{K}_2,...\) means that their joint spectrum
is simple (non-degenerated). Let \(P_1,P_2,...\) be orthogonal
projectors in \(\mathcal{H}\) projecting to the subspaces of
subspace  \({\rm Im \mit} \hat{E}_0\) (i.e.
\(P_i\hat{E}_0=\hat{E}_0P_i=P_i\)) and are projections to their own
subspaces of operators family \(\hat{H},\hat{K}_1,\hat{K}_2,...\).
All \(P_1,P_2,...\) are clearly one-dimensional due to completeness
of operators family \(\hat{H},\hat{K}_1,\hat{K}_2,...\). The
generalized microcanonical distribution (Prokhorenko and Matveev,
2011) describing our system, can be taken, for example, in the
following form:
\begin{eqnarray}
\rho=P_f, \label{1}
\end{eqnarray}
for any \(f\).

Observed values of the integrals \(K_i\), \(i=1,2,...\) in the state
\(\rho\) are given by the following formula:
\begin{eqnarray}
K'_i={\rm tr \mit}(\rho K_i).
\end{eqnarray}
Let \(SO(3)\) be a group of self-rotations in Euclidean
three-dimensional space, \(o\) is an arbitrary element of this
group, \(\hat{o}\) is a unitary representation of \(o\) in the state
space \(\mathcal{H}\) of our system. When subjected to
transformation \(o \in SO(3)\) the  state \(\rho\) comes to
\(\hat{o}\rho\hat{o}^+)\). Let's show that, if necessary, replacing
the complete set \(\hat{H},\hat{K}_1,\hat{K}_2,...\) of observables
by another complete set \(\hat{H},\hat{L}_1,\hat{L}_2,...\) allows
us to choose \(f\) from (\ref{1}) so that for some \(o \in SO(3)\)
and for some integer \(i\)
\begin{eqnarray}
{\rm tr \mit}(\rho K_i)\neq{\rm tr \mit}(\hat{o}\rho\hat{o}^+ K_i).
\end{eqnarray}
If the stated conclusion is false, then \(\forall i,j=1,2...\),
\(\forall o \in SO(3)\) , we have
\begin{eqnarray}
{\rm tr \mit}(\hat{o}P_i\hat{o}^+P_j)={\rm tr
\mit}(P_iP_j)=\delta_{ij}.
\end{eqnarray}
But the latest means that \(\forall i=1,2,...\)
\begin{eqnarray}
\hat{o}P_i\hat{o}^+=P_i. \label{2}
\end{eqnarray}
Let \(f_i\) be a unitary vector stretching the image \(P_i\). It
follows from (\ref{2}) that \(\forall i=1,2...\)
\begin{eqnarray}
\hat{o}f_i={\rm exp \mit}(i\varphi_i(o))f_i \label{3}
\end{eqnarray}
for some functions  \(\varphi_i(o)\) on \(SO(3)\). The last
conclusion is true for any complete set of observables
\(\hat{H},\hat{K}_1,\hat{K}_2,...\). In particular it is clear that
in (\ref{3}) we can choose arbitrary the orthonormal basis
\(\{f_i\}\) in \({\rm Im \mit} \hat{E}_0\). Thus \(\forall o \in
SO(3)\), the restriction of \(\hat{o}\) on \({\rm Im \mit}
\hat{E}_0\) should be diagonal in any orthonormal basis, therefore,
the restriction of \(\hat{o}\) on \({\rm Im \mit} \hat{E}_0\) must
be proportional to identical operator. But \(SO(3)\) has no
one-dimensional representations except the trivial one. Therefore,
\(\forall o \in SO(3)\) \(\hat{o}=1\). Thus, any ground state of our
system under each rotation should come to itself; but it is false as
we seen above. The statement is established.

So, the generalized microcanonical distribution \(P_f\) have the
property that some rotation of a system causes the change of
integral  \(K_i\) averaged over this state for some \(f\),
\(i=1,2,...\).

Now, if we give a sufficiently small non-zero temperature to our
system, then (as it follows from the principle of physical
continuity) for this temperature there exists a (generalized)
equilibrium state described by a generalized microcanonical
distribution \(\rho\), such that after some system rotation the
corresponding observable value of integral \(K_i\) must change for
some \(i = 1,2...\). But the system entropy does not change under
the rotation of the system. This means that for a fixed energy
(corresponding to enough small temperatures) the system entropy has
a plateau of \(d > 0\) dimensions that provides the stability of our
generalized microcanonical distributions, as it was discussed in our
previous work (Prokhorenko and Matveev, 2011).

In addition, let's remark that instead of speaking about the
completeness of the system of observables
\(\hat{H},\hat{K}_1,\hat{K}_2,...\), we should speak about the
macroscopical completeness of this system. We say that the system of
macroscopical observables quantities \(\hat{O}_1,\;\hat{O}_2,...\)
is macroscopically complete if any macroscopical quantity (which
clearly commutates with \(\hat{O}_1,\;\hat{O}_2,...\) because all
macroscopical quantities are simultaneously measurable) is a
function of observables quantities \(\hat{O}_1,\;\hat{O}_2,...\).

Thus, for enough small temperatures the system have stable
stationary states which are described by generalized microcanonical
distributions which are not reduced to the common microcanonical
distribution. We identify the crystal states of the matter exactly
with such states, and the stability of the crystal state could be
explained by just proven stability of the corresponding
microcanonical distribution.

\section{Van der Waals Model of Protoplasm} The general description
of this model is presented in the introduction. Within the framework
of this model and basing on our generalized thermodynamics, we give
numerical estimates for some changes proceeding in a cell while it
excites or becomes damaged: the amount of emitted heat by the cell
and amount of released potassium ions from the cell to environment
(according to Ling, potassium ions in the resting state are bounded
by proteins).

Let's consider two extreme protoplasm states: resting state and
"dead" protoplasm. First, let's discuss how does the dead protoplasm
appear in the context of our model.

In the dead protoplasm protein molecules are in the folded state
Prokhorenko and Matveev, 2011), and we suppose they are homogeneous
balls of radius \(r_0\) and dielectric permittivity
\(\varepsilon'\). The dielectric permittivity of the other matter in
the cell is denoted by \(\varepsilon\). Assume that \(M\) is a mass
of a protein molecule. Since a protein molecule contains a lot of
atoms, \(M\) is a very large quantity.

Let's write out the equation of motion which describes the motion of
protein molecules. We suppose that the cell is described by
classical mechanics, however further obtained results clearly imply
that the answer for the quantum case is the same. The motion of
protein molecules may be considered as classical motion because of a
high value of \(M\). Let's denote \(x=(p,q)\) are coordinates and
momenta of all protein molecules. Put by definition that
\(y=(p',q')\) are coordinates and momenta of all other protoplasm
components. Let \(H(x,y)\) be the Hamiltonian of the whole
protoplasm. Then Hamilton's differential equations on \(x\) take the
following form:
\begin{eqnarray}
\dot{p}=-\frac{\partial H(x,y)}{\partial q},\nonumber\\
\dot{q}=\frac{\partial H(x,y)}{\partial p} \label{I1}.
\end{eqnarray}
However, since the cell is dead, its distribution function is a
Gibbs distribution function. In particular, the conditional
probability density that variable \(x\) takes value \(x'\) provided
variable \(y\) takes value \(y'\) is given by:
\begin{eqnarray}
w(x'|y')=\frac{1}{Z_1(x')}e^{-\frac{H(x',y')}{T}},
\end{eqnarray}
where \(T\) is a system temperature and
\begin{eqnarray}
Z_1(y'):=\int d x e^{-\frac{H(x,y')}{T}}.
\end{eqnarray}

Since protein molecules move slowly and their mass \(M\) is very
large (thousands of \(D\)), in (\ref{I1}) we can replace right parts
by their averages over distribution \(w(y|x)\). Omitting rather
trivial calculations, we find that the averaged system (\ref{I1}) is
a Hamiltonian one too and the corresponding Hamiltonian is a free
energy of the system. More specifically, the averaged system
(\ref{I1}) is given by:
\begin{eqnarray}
\dot{p}=-\frac{\partial F(x|T)}{\partial q},\nonumber\\
\dot{q}=\frac{\partial F(x|T)}{\partial p} \label{I2},
\end{eqnarray}
where
\begin{eqnarray}
F(x|T):=-T \ln \int d y e^{-\frac{H(x,y)}{T}}.
\end{eqnarray}
This is a standard adiabatic limit.

Note two more properties.

1.  If our whole system (protoplasm) is described by a Gibbs
distribution:
\begin{eqnarray}
w(x,y)=\frac{1}{Z} e^{-\frac{H(x,y)}{T}} \label{I12},
\end{eqnarray}
then, the distribution of probability that protein molecules are
situated in the given point of phase space can be achieved by
integrating (\ref{I12}) by \(dy\). The distribution of probability
\(w(x)\) to find protein molecules in the given point of the
configuration space is:
\begin{eqnarray}
w(x)={\rm const \mit} e^{-\frac{F(x|T)}{T}}.
\end{eqnarray}
That is, again we received the Gibbs distribution in which the
Hamiltonian is the effective Hamiltonian for protein molecules we
received above.

2.  By using a standard formula we can calculate the free energy of
protein system \(F'(T)\) for Hamiltonian  \(F(x|T)\). Elementary
calculations give:
\begin{eqnarray}
F'(T)=F(T):=-T \ln \int dx dy e^{-\frac{H(x,y)}{T}},
\end{eqnarray}
i.e. \(F'(T)\) equals to the free energy of the whole system
\(F(T)\).

Now let's define the form of our effective Hamiltonian \(F(x|T)\) as
a function of coordinates and momentas of protein molecules. We
suppose the contribution to \(F(x|T)\) nontrivially dependent on
\(x\) is caused by Van der Waals interaction between protein
molecules and therefore \(F(x|T)\) is given by:
\begin{eqnarray}
F(x|T)=E_{kin}(p)+F_0(T)+\sum \limits_{i>j} V(q_i-q_j|T),
\end{eqnarray}
where \(E_{kin}(p)\) is a kinetic energy of protein molecules as
material points, \(F_0(T)\) is a function of temperature and
\(q_i\;p_i\), \(i = 1,2,3,...\) are Cartesian coordinates of protein
molecules and canonically conjugated momenta. \(V(x|T)\)--- is a
pair interaction potential of the following form:
\begin{eqnarray}
V(q|T)=+\infty\;{\rm if \mit}\;|q|<2r_0,\nonumber\\
V(q|T)=-\frac{C}{r^6},\;{\rm if \mit}\;|q|\geq 2r_0,
\label{POTENTIAL}
\end{eqnarray}
where \(C\) is some positive constant. The explicit formula
expressing \(C\) in terms of \(\varepsilon\), \(\varepsilon'\),
\(r_0\) can be retrieved, for example, from (Lifshitz and Pitaevsky,
1978). Now, we find thermodynamic variables for the case just
described. The solution of this problem providing that
\begin{eqnarray}
\frac{\min_{2r_0<r}|V(r|T)|}{T}\ll1 \label{LIP}
\end{eqnarray}
is stated in many textbooks, for example (Landau, Lifshitz, 1995).
However, in real cells (for example, erythrocyte) this condition is
not fulfilled; this is a subject of a special analysis in the next
section. For simplicity here we suppose that condition (\ref{LIP})
is fulfilled.

In the case of the living cell, as we shall see later, the free
energy of a cell is given by expression:

\begin{eqnarray}
F(V,T)_l=F_0(V,T)+F_{id}(V,T), \label{SS1}
\end{eqnarray}

where  \(F_0(V,T)\) is a free water energy where all the proteins
are eliminated and \(F_{id}\) is free protein energy calculated as
if it was an ideal gas. For the dead protoplasm
\begin{eqnarray}
F(V,T)_l=F_0(V,T)+F_{id}(V,T)+\Delta F(V,T), \label{SS2}
\end{eqnarray}
where
\begin{eqnarray}
\Delta F(V,T)=-T \ln [ \int \limits_V ...\int \limits_V \frac{d^3
q_1}{V}....\frac{d^3 q_N}{V} e^{-\frac{1}{T} \sum \limits_{1\leq
i<j\leq N} V(q_i-q_j|T)}]. \label{CONF}
\end{eqnarray}
Note that when we use formulas (\ref{SS1}), (\ref{SS2}),
(\ref{CONF}), we neglect the conformational part of the free energy.
There is one more omitted contribution to the free energy, which is
caused by possibility of protein molecules rotation as a unit.
However, these contributions for fully unfolded and folded protein
conformation differ only by the value of \({\rm const\; \mit} T\) and
therefore, as it will be clear from the following, omitting of those
contributions does not influence the final results.

For the case when (\ref{LIP}) is fulfilled and a gas is so much
rarefied that we can take into account only pair collisions, the
calculations of integral (\ref{CONF}) are made in many textbooks
(for example, see (Landau and Lifshitz, 1995). But to find \(\Delta
F(V,T)\) in the real case we should derive an expression for
\(\Delta F(V,T)\) when condition (\ref{LIP}) is fulfilled and only
pair collisions are taken into account. Therefore, we give here
derivation for \(\Delta F(V,T)\) (condition (\ref{LIP}) is
fulfilled). Put by definition \((U(q|T):= \sum \limits_{1\leq
i<j\leq N} V(q_i-q_j|T))\). Then
\begin{eqnarray}
\Delta F(V,T)=-T \ln [ \int \limits_V ...\int \limits_V \frac{d^3
q_1}{V}....\frac{d^3 q_N}{V} \{e^{-\frac{U(q|T)}{T}}-1\}+1]
\end{eqnarray}
If we take into account only pair collisions and suppose them rare,
then the whole configuration space of the system
\(\mathcal{C}=\mathbb{R}^3\times...\times \mathbb{R}^3 \) should be
divided into subareas with equal volume \(\mathcal{C}_{i,j}\),
\(1\leq i<j\leq N\) such that in each of them collisions of \(i\)-th
and \(j\)-th particles happens. But in \(\mathcal{C}_{i,j}\)
\(U(q|T)=V(q_i-q_j|T)\) So:
\begin{eqnarray}
\Delta F(V,T)=-T \ln [ \sum \limits_{1\leq i<j\leq N} \int
\limits_{\mathcal{C}_{i,j}}  \frac{d^3 q_1}{V}....\frac{d^3 q_N}{V}
\{e^{-\frac{V(q_i-q_j|T)}{T}}-1\}+1].
\end{eqnarray}
But pairs \((i,j)\) such, that \(1\leq i<j\leq N\) could be chosen
by \(\frac{N(N-1)}{2}\approx \frac{N^2}{2}\) ways. So, we have

\begin{eqnarray}
\Delta F(V,T)=-T \ln [ \frac{N^2}{2V^2}  \int \limits_{V} \int
\limits_{V} d^3 q_1 d^3 q_2
\{e^{-\frac{V(q_1-q_1|T)}{T}}-1\}+1]=\nonumber\\
-T \ln [ \frac{N^2}{2V}  \int \limits_{V} d^3 q
\{e^{-\frac{V(q|T)}{T}}-1\}+1]
\end{eqnarray}

Expanding the logarithm in a Taylor series near \(1\), the final
result is

\begin{eqnarray}
\Delta F(V,T)=-T \frac{N^2}{2V}  \int \limits_{V} d^3 q
\{e^{-\frac{V(q|T)}{T}}-1\}.
\end{eqnarray}

In the area \(r:=|q|<2r_0\) \(\{e^{-\frac{V(q|T)}{T}}-1\}=-1\) and
in the area \(r>2r_0\)
\(\{e^{-\frac{V(q|T)}{T}}-1\}=-\frac{V(q|T)}{T}\) due to weakness of
interaction. Put by definition:
\begin{eqnarray}
b=\frac{16 {\pi}r_0^3}{3},\nonumber\\
a=2\pi \int \limits_{2r_0}^{+\infty} |V(r|q)|r^2dr.
\end{eqnarray}
Notice by the way that using our choice of potential
\begin{eqnarray}
a=\frac{\pi C}{12r_0^3}.
\end{eqnarray}
With these notations we can rewrite \(F(V,T)\) in the following way:
\begin{eqnarray}
\Delta F(V,T)= T \frac{N^2}{V}(b-\frac{a}{T}).
\end{eqnarray}
So
\begin{eqnarray}
S_d=S_0+S_{id}-b\frac{N^2}{V},\nonumber\\
E_d=E_0+E_{id}-\frac{N^2a}{V},
\end{eqnarray}
where \(S_d\) and \(E_d\) are entropy and energy of the dead
protoplasm respectively, \(S_0\) and \(E_0\) are entropy and energy
of the protoplasm where all the protein molecules are eliminated and
\(S_{id}\) and \(E_{id}\) are entropy and energy of the ideal gas of
protein molecules.

That is a definition of the thermodynamic features of the dead
protoplasm in our model. Now, let's consider the living protoplasm.
Let's divide the effective Hamiltonian of the protein system
obtained by the method of adiabatic limit into two summands:
\begin{eqnarray}
H=H_{v-d-W}+H',
\end{eqnarray}
Here \(H_{v-d-W}\) contains the kinetic energy of proteins as
material points and energy of the Van der Waals interaction between
them. The Hamiltonian \(H'\) depends on variables describing
internal degrees of freedom of proteins. We suppose that in a
certain sense \(H'\ll H_{k}\), where \(H_k\) is the kinetic energy
of all protein molecules. Next, according to our common view
(Prokhorenko and Matveev, 2011), in the living protoplasm some first
integrals in the involution \(K_1,...,K_n\) are active and a
statistical weight of protein molecules is given by:
\begin{eqnarray}
W(E)=\int \prod \limits_{i=1}^N dp_idq_i \prod
\limits_{j=1}^n\delta(K_j-K'_j)\delta(H_{v-d-V}+H'-E), \label{J2}
\end{eqnarray}
where \(p_i\), \(q_i\) are momenta and coordinates of \(i\)-th
protein.

Let's try to define the form of integrals \(K_i\). According to
Ling, intracellular water in the resting living protoplasm is in the
bound state and protein molecules form a paracrystals. Since we
suppose the living state differs from non-living by activity of
integrals \(K_i\), then it's reasonable to assume that fixation of
protein molecules in points of lattice is performed by means of
multiplier \(\prod \limits_{j=1}^n\delta(K_j-K'_j)\) in the integral
of the statistical weight definition. Therefore, we just suppose
that
\begin{eqnarray}
\prod \limits_{j=1}^n\delta(K_j-K'_j)=\prod
\limits_{i=1}^N\delta(q_i-q'_i),
\end{eqnarray}
where \(q_i\) are coordinates of \(i\)-th point of protein molecules
lattice. But on the support of \(\prod
\limits_{i=1}^N\delta(q_i-q'_i)\) the potential energy of protein
interaction \(\sum \limits_{i>j} V(x_i-x_j|T)\) is a constant.
Furthermore, we can suppose that on the support \(\prod
\limits_{i=1}^N\delta(q_i-q'_i)\) the potential energy of proteins
\(\sum \limits_{i>j} V(q_i-q_j|T)=0\). Indeed, if proteins are
situated in points of the lattice we mentioned above, then
\begin{eqnarray}
\sum \limits_{i>j} V(q_i-q_j|T)\sim \frac{N}{r^6},
\end{eqnarray}
where \(r\) is a minimal distance between proteins. But
\(r\sim(\frac{V}{N})^{1/3}\). Therefore:
\begin{eqnarray}
\sum \limits_{i>j} V(q_i-q_j|T)\sim N(\frac{N}{V})^{2}. \label{33}
\end{eqnarray}

Further we show that if the potential energy of protein molecules
interaction is neglected, then, when the cell is dying, the quantity
of emitted heat and the number of potassium ions released from the
cell, calculated for one protein molecule, is a linear polynomial in
\(\frac{N}{V}\) with an accuracy up to logarithmic factors. So, as
follows from (\ref{33}), in (\ref{J2}) in the limit of low density
the potential energy of protein interaction in \(H_{v-d-W}\) can be
neglected just kinetic energy \(H_k\). Eventually, for the
statistical weight:
\begin{eqnarray}
W(E)=\int \prod \limits_{i=1}^N dp_idq_i \prod
\limits_{j=1}^N\delta(q_j-q'_j)\delta(H_{k}+H'-E).
\end{eqnarray}
Since \(H'\ll H_{k}\) we can write over this expression in the
following way:
\begin{eqnarray}
W(E)=\int \prod \limits_{i=1}^N
 dp_idq_i \prod \limits_{j=1}^N
\delta(q_j-q'_j)\delta(\sum \limits_{i=1}^N \frac{p_i^2}{2M}-E).
\end{eqnarray}
Consequently with an accuracy of inessential multiplier, the
statistical weight \(W\) equals to the statistical weight for the
ideal gas. Therefore, in our model thermodynamic variables for the
living protoplasm take the form of:
\begin{eqnarray}
E_l=E_0+E_{id},\nonumber \\
S_l=S_0+S_{id}.
\end{eqnarray}
Here \(E_l\) and \(S_l\) are entropy and energy of the living
protoplasm. Obviously \(E_d< E_l\), i.e. our model predicts that
activation and death of the protoplasm is an exothermal reactions.
Numerical evaluations are given in the next section. So, we have the
following amount for the quantity of released heat:
\begin{eqnarray}
Q=\frac{N^2a}{V}.
\end{eqnarray}
The ability of variables  \(q_i\) or closed to them play the role of
motion integrals in our model is appeared from the following
additional conclusions. Let's denote by \(E_{kin}\) a kinetic energy
of one protein molecule. Obviously
\begin{eqnarray}
|\dot{q}_i|\leq \sqrt{\frac{2E_{kin}}{M}}\sim \sqrt{\frac{T}{M}}.
\end{eqnarray}

Therefore, in the limit \(M\rightarrow \infty\) (very large mass of
protein) \(\dot{q}_i=0\) and \(q_i\) are motion integrals.

Now we define the number of potassium ions releasing from the dying
cell. In our previous work (Prokhorenko and Matveev, 2011) the
following formula was obtained. Let's suggest that the number of
first integrals in the involution is so large that the number of
active integrals can be characterized by continuous parameter \(s
\in[0,1]\). In addition, the number of active first integrals is
increasing function of \(s\) and a case of \(s = 0\) corresponds to
the case when none of integrals is active, and case of \(s = 1\)
corresponds to the case when all the first integrals are active.
Supposes \(s\) infinitesimally varies \(s\mapsto s'=s-\delta s\),
\(\delta s>0\), where \(s\) is infinitely small. Now, let's define
the function of entropy \((\delta f)(S)\) by the condition \((\delta
f)(S)=(\delta S)_E\), ( \(s\mapsto s'=s-\delta s\)). This definition
is correct due to a proven assertion (Prokhorenko and Matveev, 2011)
that \((\delta S)_E\) ( \(s\mapsto s'=s-\delta s\), \(\delta s>0\))
is a constant along any adiabatic process. So, let  \(s\mapsto
s'=s-\delta s\), \(\delta s>0\), \(\delta s\),  is infinitely small.
As it was shown in (Prokhorenko and Matveev, 2011), the increasing
of number of potassium ions in the cell \(\delta N\) can be found
using the following formula:
\begin{eqnarray}
\delta N=T(\delta f)'(S)(\frac{\partial S}{\partial \mu})_T,
\end{eqnarray}
where \(\mu\) is a chemical potential of potassium ions in the cell.

First, let's define \((\frac{\partial S}{\partial \mu})_T\). Suppose
by definition \(\tilde{F}_0(V,T)\) is the free energy of the
protoplasm where potassium ions absent.

Suppose by definition \(\tilde{F}_0(V,T)\) is the free energy of the
ideal gas at temperature \(T\) while its mass equals to the mass of
potassium ion \(M\) and the gas contains only one molecule. It is
known (Landau and Lifshitz, 1995) that
\begin{eqnarray}
\phi(V,T)=-T \ln [V(\frac{MT}{2\pi\hbar^2})^{\frac{3}{2}}].
\end{eqnarray}
Suppose by definition \(\psi(V,T)\) is potassium solvability in the
protoplasm, i.e. change of the free protoplasm energy during
transition of one potassium ion from the infinity to the given
point. Then the change of the free protoplasm energy after adding
one potassium ion to this protoplasm is just a sum of
\(\varphi(T,V)\) and \(\psi(V,T)\). So, the free energy of all the
protoplasm is
\begin{eqnarray}
\tilde{F}_0(V,T) \mapsto
\tilde{F}(V,T)=\tilde{F}_0(V,T)+\varphi(T,V)+\psi(V,T).
\end{eqnarray}
If there are \(N\) potassium ions inserted to the protoplasm, but
\(N\) is still very small, then an interaction between different
potassium ions can be neglected and we have:
\begin{eqnarray}
\tilde{F}_0(V,T) \mapsto
\tilde{F}(V,T)=\tilde{F}_0(V,T)+N\varphi(T,V)+N\psi(V,T)+NT
\ln(\frac{N}{e}).
\end{eqnarray}
The last summand arises because of a common combinatorial multiplier
\(\frac{1}{N!}\) included to definition of the partition function of
potassium ions.

We neglect the solvability of potassium ions in our calculations, in
other words we accept \(\psi(V,T)=0\).

Put by definition
\begin{eqnarray}
\mathcal{A}(V,T):=V(\frac{MT}{2\pi\hbar^2})^{\frac{3}{2}}.
\end{eqnarray}

Eventually, after introduction of \(N\) potassium ions into the
protoplasm, the free protoplasm energy changes as follows:
\begin{eqnarray}
\tilde{F}_0(V,T) \mapsto
\tilde{F}(V,T)=\tilde{F}_0(V,T)+N\psi(T,V)+NT
\ln(\frac{N}{e\mathcal{A}(V,T)}).
\end{eqnarray}
Notice that the quantity \(\frac{N}{\mathcal{A}(V,T)}\ll 1\) since
by order of magnitude this value is a density of particles in
phase-space cells of constant-energy surface and thus it is very
small, because the gas of potassium ions, as we admitted, is
strongly rarefied (particularly, this allows us to use formulas of
the classic statistical mechanics).

Potassium adsorption is taken into account in the solvability
\(\psi(V,T)\). We supposed that this solvability equals to zero but
this assumption is not quite right. After cell death the chemical
potential of potassium ions, obtained by differentiating the free
energy over with number of potassium ions, remains constant. Since
while the cell is dying, it issues \(\approx 0.98\) of the total
amount of potassium ions, solvability \(\psi(V,T)\) increases by
\(T\ln 50\). Though this section contains exact equalities, actually
this equalities are approximate because we neglect a unity compared
to \(\ln(\frac{N}{e\mathcal{A}(V,T)})\) and change of \(\psi(V,T)\)
by value of \(T\ln 50\). However, the following analysis establishes
that contribution of those omitted summands is inessential.

When \(N\) potassium ions are inserted into the protoplasm,the
entropy changes as follows:
\begin{eqnarray}
\tilde{S}_0(V,T)\mapsto\tilde{S}(V,T)=\tilde{S}_0(V,T)-N\ln(\frac{N}{e\mathcal{A}(V,T)}).
\end{eqnarray}
We have
\begin{eqnarray}
(\frac{\partial \tilde{S}}{\partial \mu})_T=\frac{(\frac{\partial
S}{\partial N})_T}{(\frac{\partial \mu}{\partial N})_T}.
\end{eqnarray}
Further:
\begin{eqnarray}
(\frac{\partial \tilde{S}}{\partial
N})_T=-\ln(\frac{N}{e\mathcal{A}(V,T)}).
\end{eqnarray}
And:
\begin{eqnarray}
\mu=(\frac{\partial\tilde{F}(V,T,N)}{\partial
N})_T=T\ln(\frac{N}{e\mathcal{A}(V,T)}).
\end{eqnarray}
Therefore:
\begin{eqnarray}
(\frac{\partial \mu}{\partial N})_T=\frac{T}{N}.
\end{eqnarray}
Eventually:
\begin{eqnarray}
(\frac{\partial \tilde{S}}{\partial
\mu})_T=\frac{N}{T}\ln[\frac{e\mathcal{A}(V,T)}{N}],
\end{eqnarray}
and
\begin{eqnarray}
\delta N=N(\delta
f)'(S)\ln[\frac{e\mathcal{A}(V,T)}{N}].\label{DIFFUR}
\end{eqnarray}
\section{Van der Waals Model of Protoplasm. Numerical Evaluations 1. Potassium Ions Efflux from the Cell and Heat Release}
This section is concerned with getting numerical evaluations basing
on the Van der Waals model (see above) and comparing them with
experimental data. To specify model parameters (cell size, quantity
of proteins and ions in the cell) we chose a human erythrocyte, the
well-studied cell having a relatively simple structure-function
organization.

It's known that potassium ions density in the living erythrocyte is
estimated as \(n=6.02\times10^{19} cm^{-3}\). Let's assume that when
an erythrocyte is dying the potassium concentration in this
erythrocyte becomes equal to the potassium concentration in the
blood plasma, \(2-4\, \rm mmol/l\mit\), i.e. about \(0.96-0.98\) of
potassium ions, which the living cell contained, release from the
erythrocyte.

We consider the cell at a temperature of 300\(K\). A potassium
nucleus contains 19 protons and 20 neutrons, therefore, the mass of
potassium ions can be estimated as \(39 m_n\) where \(m_n\) is the
mass of neutron.

In the previous section the formula was derived describing the
change of a number of potassium ions in the cell when the
infinitesimal change of a number of active integrals \(s\mapsto
s'=s-\delta s\), \(\delta s\) is infinitely small. Let's write it
one more time:
\begin{eqnarray}
\delta N=N(\delta f)'(S)\ln[\frac{e\mathcal{A}(V,T)}{N}].
\end{eqnarray}
The last formula (\ref{DIFFUR}) can be interpreted as a differential
equation on \(N\). Being integrated this equation expresses of
potassium efflux from the dying cell through other cell parameters.

First, let's find \((\delta f)'(S)\). We have:
\begin{eqnarray}
\delta f(S)=(\delta S)_E=(\delta S)_T-(\delta E)_T (\frac{\partial
S}{\partial E})_T.
\end{eqnarray}
But
\begin{eqnarray}
(\frac{\partial S}{\partial E})_N=\frac{1}{T}
\end{eqnarray}
Therefore
\begin{eqnarray}
\delta f(S)=(\delta S)_E-\frac{1}{T} (\delta E)_T.
\end{eqnarray}
Using the fact that (see Prokhorenko and Matveev, 2011)
\begin{eqnarray}
\frac{\partial (\delta S)_T}{\partial (\delta E)_T}=\frac{1}{T},
\end{eqnarray}
we find
\begin{eqnarray}
\frac{d \delta f(S(T))}{d T}=\frac{1}{T^2} (\delta E)_T.
\end{eqnarray}
And finally
\begin{eqnarray}
(\delta f)'(S)=\frac{1}{T^2} (\delta E)_T\frac{\partial T}{\partial
S} =\frac{1}{C_V T}(\delta E)_T,
\end{eqnarray}
where \(C_V\) is the heat capacity of the cell at the constant
volume. Now, we assume the heat capacities of living and dead cells
are almost the same, and in all following formulas the heat capacity
of a cell can be replaced by an average value \(C_V\) which does not
depend on the number of active first integrals in the involution.
Eventually, we have
\begin{eqnarray}
\frac{\delta N}{N}=-\frac{1}{TC_V}(\delta E)_T
\ln[\frac{N}{e\mathcal{A}(V,T)}].
\end{eqnarray}

For convenience of further calculations let's introduce a new
variable
\begin{eqnarray}
\Lambda:=\frac{N}{e\mathcal{A}(V,T)}.
\end{eqnarray}

Then we have:
\begin{eqnarray}
\delta \ln \Lambda=-\frac{1}{TC_V} (\ln \Lambda)(\delta E)_T.
\label{DIFFUR1}
\end{eqnarray}

As before, we use lower indexes \(l\) and \(d\) to denote variables
relating to the living and dead cell respectively. Equation
(\ref{DIFFUR1}) is easy to integrate resulting:
\begin{eqnarray}
\frac{\ln \Lambda_d}{\ln \Lambda_l}={\rm exp \mit} (\frac{Q}{C_VT}),
\label{FIRST}
\end{eqnarray}
where \(Q\) is a heat amount released from the cell while it is
dying.

The last formula gives an implicit expression for heat generation
\(Q\) and its derivation was based just on common thermodynamic
considerations without regard to properties of any certain model.
Therefore, it can be used for verification of our generalized
thermodynamics.

However, note that (\ref{FIRST}) is inconvenient for calculations,
so, let's simplify it using smallness of \(\Lambda_l\). For this
purpose let's consider the expression \(\ln x\), where \(x\) is a
very small positive number. Then \(\ln x\) value is very large in
modulus and has a sign minus. Let's increase \(x\) by factor of
\(k\) where \(k\) is not too large natural number \(\ln x \mapsto
\ln x +\ln k\) i.e. practically unalters. We have:
\begin{eqnarray}
\frac{\ln \Lambda_d}{\ln \Lambda_l}=|\frac{\ln \Lambda_d}{\ln
\Lambda_l}|={\rm exp \mit}\{ \int \limits_{|\ln \Lambda_l|}^{|\ln
\Lambda_d|} \frac{dx}{x}\}.
\end{eqnarray}
According to the newly stated remark, on the whole integrating
interval we can replace \(x\) by \(|\ln \Lambda_l|\) and eventually
we obtain:
\begin{eqnarray}
\frac{\ln \Lambda_d}{\ln
\Lambda_l}=\{\frac{\Lambda_l}{\Lambda_d}\}^{\frac{1}{|\ln
\Lambda_l|}}.
\end{eqnarray}
Eventually, we receive the following implicit expression for heat
generation:
\begin{eqnarray}
\frac{N_l}{N_d}={\rm exp \mit} \{|\ln \Lambda_l| \frac{Q}{C_VT}\}.
\end{eqnarray}
Data listed in this section are enough to calculate \(|\ln
\Lambda_l|\). Omitting corresponding numerical calculations, we
present the result: \(|\ln \Lambda_l|\approx 16.1\). The resulting
formula for heat generation is:
\begin{eqnarray}
\frac{N_l}{N_d}={\rm exp \mit}\{\frac{16.1Q}{C_VT}\}.\label{STAR}
\end{eqnarray}
From here it is easy to find \(\frac{Q}{C_VT}\).
\(\frac{N_l}{N_d}\approx50\). \(\ln 50=3.93\). By taking a logarithm
of both parts we find:
\begin{eqnarray}
\frac{Q}{C_VT}\approx 0.24.
\end{eqnarray}

To check this heat generation value we can carry the following
qualitative reasoning. Let's consider a human cell at temperature
\(T_h=37 C=310 K\) or 37\({}^0C\). The room temperature (normal for a
man's functioning) \(T_r=20 C=293 K\). According to Ling's theory,
the cell life activity is expressed as the cycle motion
"resting\(\leftrightarrow\)exciting". If the cell was always in
thermodynamic equilibrium with the thermostat, then, at transition
"resting\(\leftrightarrow\)exciting", the \(Q\) heat would released.
Let's accept that in the resting state a cell has a room
temperature. Let \(a\) be a digit making \(Q=\alpha C_VT\). If the
cell is heat sealed, then in transition from the resting state to
the excited state the temperature of the cell rises from \(T_r\) to
\(T_e\) and it is likely to be correct that there is an approximate
correlation \((T_e-T_r)/T_h=\alpha\). Now, as the cell moves through
time cyclically, we have a mean temperature for time \(T_h\). It is
reasonable to suggest that \(T_h=(T_e+T_r)/2\), or in other words
\(T_e-T_r=2(T_h-T_r)\). But \(T_h-T_r=17 K\). Therefore
\(T_e-T_r=34K\) and \(\alpha=(T_e-T_r)/T_h=34/310=0.11\), which
coincides with our result in the order of magnitude. The fact that
the calculation for heat emission of the dying erythrocyte using
data on potassium efflux from the cell appears approximately 2.5
times higher than just stated value is reasonable because the
exciting can be considered as a stage towards death (Matveev, 2005).

The assumption we stated that the erythrocyte temperature at rest
equals to the room one \(T_{room}\) can be explained in the
following way. Since the erythrocyte moves through time cyclically:
resting\(\leftrightarrow\)exciting, then the erythrocyte can be
considered as a heat engine, and room temperature is a temperature
of a cooler of this engine. As the room temperature is the most
comfortable for a man, we may consider that the erythrocyte at the
room temperature, as at the cooler temperature, operates in the most
optimum mode. The resting temperature \(T_r\) is the lowest
temperature achieved by the erythrocyte during all the resting
 \(\leftrightarrow\) exciting cycle. If \(T_r>T_{room}\) is correct,
then heat transfer from the erythrocyte to the cooler will beat
tended by heat transfer from a warmer body to a colder one,
i.e. entropy increase. This means the erythrocyte as a heat engine
would operate not optimally. Conversely suppose that
\(T_r<T_{room}\). If the erythrocyte is functioning in the optimum
mode (without entropy increase), then the erythrocyte have to pass
the part of the cycle when the erythrocyte temperature \(T <T_r\)
being surrounded by an adiabatic "cover". It becomes
incomprehensible why does the erythrocyte need this part of the
cycle and also, according to Carnot's theorem on the efficiency of
heat engines, erythrocyte efficiency could be raised by means of the
cooler temperature (room temperature) decreasing.

Now let's try to calculate \(Q\) basing on our investigated (Van der
Waals) model. For this purpose we should find an interaction
constant \(C\) in the law \(V(r)=\frac{C}{r^6}\). To do this, we
proceed from the formula taken from (Lifshitz and Pitaevsky, 1978).
Suppose there are two parallel non-overlapping semi-spaces and a
distance between them is \(l\). Let \(\varepsilon\) be a dielectric
permittivity of the semi-spaces and  \(\varepsilon'\) is a
dielectric permittivity of the cavity between them. Suppose \(l\) is
such a large that if \(\omega\) is a frequency of the electromagnetic
wave distributed in the cavity between semi-spaces and having
wavelength \(l\), then \(\hbar\omega \ll T\). That is obviously our
case. Then, the force of attraction between two semi-spaces as per
unit area of the border space of each semi-space is equal to:
\begin{eqnarray}
P=\frac{T}{16 \pi^2 l^3} \int \limits_0^{+\infty}
x^2[\{\frac{\varepsilon+\varepsilon'}{\varepsilon-\varepsilon'}\}^2e^x-1]^{-1}dx.
\end{eqnarray}
Now, suppose that the semi-spaces consist of folded hemoglobin
molecules and the cavity between them is filled with water. The
dielectric permittivity of the water \(\varepsilon=81\) and the
dielectric permittivity of the hemoglobin \(\varepsilon'\approx 2\).
Therefore, the factor
\(\{\frac{\varepsilon+\varepsilon'}{\varepsilon-\varepsilon'}\}^2\approx1\)
and the force of attraction between two semi-spaces as per unit area
of the border space of each of them is equal to
\begin{eqnarray}
P=\frac{T}{16 \pi l^3} \int \limits_0^{+\infty} x^2[e^x-1]^{-1}dx.
\end{eqnarray}
The involved integral can be easily calculated with any desired
degree of precision, for example:
\begin{eqnarray}
\int \limits_0^{+\infty} x^2[e^x-1]^{-1}dx=\int \limits_0^{+\infty}
x^2e^{-x}[1-e^{-x}]^{-1}dx=\nonumber\\
=\int \limits_0^{+\infty}
x^2[e^{-x}+e^{-2x}+e^{-3x}+.....]dx=\nonumber\\
=2[1+\frac{1}{2^3}+\frac{1}{3^3}+....]=2[1+1/8+1/27+1/64+...]\approx2.35.
\end{eqnarray}
Therefore, we come to the following formula for the force of
attraction between the semi-spaces:
\begin{eqnarray}
P=\frac{2.35 T}{16 \pi l^3}. \label{KAZIMIR}
\end{eqnarray}
Let \(V_{1H}\) be a volume of one hemoglobin molecule. Now, we can
calculate a potential energy \(\mathcal{U}(l)\)  per unit of the
surface plane using a formula for attraction potential of two
hemoglobin molecules (\ref{POTENTIAL}) in the following way:
\begin{eqnarray}
\mathcal{U}(l)=-\frac{C}{V_{1H}^2} 2\pi \int \limits_0^{+\infty} r
dr \int \limits_0^{+\infty}dx \int \limits_0^{+\infty} dy
\frac{1}{(x+y+l)^2+r^2)^3}.
\end{eqnarray}
Omitting rather trivial integrating we find
\begin{eqnarray}
\mathcal{U}(l)=-\frac{\pi C}{12V^2_{1H}}\times \frac{1}{l^2}.
\end{eqnarray}
Differentiating the last expression with respect to \(l\) we find:
\begin{eqnarray}
P=\frac{\pi C}{6V^2_{1H}}\times \frac{1}{l^3}
\end{eqnarray}

Comparing the last formula with (\ref{KAZIMIR}) gives the following
result:
\begin{eqnarray}
C=V_{1H}^2 \frac{7.05 T}{8 \pi^2}.
\end{eqnarray}

It follows
\begin{eqnarray}
a=V_{1H}T\times \frac{7.05}{72}=9.79\times 10^{-2} V_{1H} T
\end{eqnarray}

Note that everywhere above we measured the temperature \(T\) in
energy units. Let \(T'\) be an absolute temperature expressed in
Kelvin degrees. There is a relation \(T=k_B T'\) where
\(k_B=1.38\times 10^{-16} erg K^{-1}\). Hereafter \(Q_v\) means the
heat released from a cell within the framework of the Van der Waals
model. We want to estimate \(\frac{Q_v}{CT}\). Then, for heat
generation:
\begin{eqnarray}
O_v=1.35\times10^{-17} N^2 (\frac{V_H}{V})(\frac{V_{1H}}{V_H})
erg\times K^{-1}T'
\end{eqnarray}
where \(V_H\) is the volume of all hemoglobin contained in the dead
erythrocyte.

We need to know the cell volume \(V\), volume of all hemoglobin
\(V_H\) contained in the dead cell, mass of all the hemoglobin
\(M_H\), specific heat capacity of hemoglobin per unit mass \(C_M\).
Calculating this data we find \(V=10^{-10}  cm^3\) (Levine et al., 2001). \(C_M=3.2\times 10^7 erg\; g^{-1} K^{-1}\) (Kholodny et al., 1987). Further it's known that hemoglobin concentration \(C_H=5 mmol \times
L^{-1}\), \(C_H=5\times 10^{-6} mol\times cm^{-3}\) (Van Beekvelt et al.,
 2001). The volume of "dead" hemoglobin
is \(V_H=32.6\times10^{3} cm^3/mol\) (Arosio et al., 2002). \(N=3.02\times10^8\) (Van Beekvelt et al., 2001). Therefore,
\begin{eqnarray}
\frac{V_H}{V}=0.163
\end{eqnarray}

But hemoglobin concentration in the human erythrocyte is \(0.33
g/cm^3\) (Van Beekvelt et al., 2001) and its density in the dead cell (in the supercluster)
is approximately two times greater than the water density (Van Beekvelt et al., 2001;
Arosio et al., 2002).
It follows the mass of the erythrocyte \(M_C=1.16\times 10^{-10}
g\). The heat capacity \(C_V\) is \(C_V=M_C C_M=3.71\times 10^{-3}\;
erg \;K^{-1}\).

Let's calculate  \(N^2\frac{V_H}{V} \). Taking into account that
\(N^2=9.12\times 10^{16}\), we find
\begin{eqnarray}
\frac{N^2 V_H}{V}=1.49\times 10^{16}.
\end{eqnarray}
Eventually
\begin{eqnarray}
\frac{Q_v}{C_V T}=54 \frac{V_{1H}}{V_H}
\end{eqnarray}
Here we see that if we take a true value \(\approx 3.3 \times
10^{-9}\)  for  \(\frac{V_{1H}}{V_H}\), then the resulting value for
\(\frac{Q_v}{C_V T}\) is many times smaller than \(\frac{Q_v}{C_V
T}=0.24\) we found before.

This fact has the following explanation. Our derivation of
corrections to the free energy was correct only upon the following
condition:
\begin{eqnarray}
\frac{\min_{2r_0<r}|V(r|T)|}{T}\ll1 \label{LIPA}
\end{eqnarray}
Let's calculate the value in the right part of this inequality.
Suppose two protein molecules are situated so that a distance
between their centers is slightly more than \(2r_0\). Then, the
potential energy between them is  \(U=-C\frac{1}{(2r_0)^6}\). We have
shown above that  \(C=V_{1H}^2 T \frac{7.05}{8 \pi^2}\). Here, after
elementary calculations, we find
\begin{eqnarray}
\frac{U}{T}=\frac{7.05}{288}\approx 2.5 \times 10^{-2},
\end{eqnarray}
i.e. our criterion is really fulfilled. However, when two protein
molecules come to each other such close that \(2 \pi\hbar
\frac{c}{l} \approx 1\), the forces whose contribution we did not
take into consideration before begin to play a significant role. We
mean Casimir forces caused by the energy of zero-point oscillations
of the electromagnetic field in the space between proteins, and
which significantly exceed forces we took into consideration before.
These forces can be interpreted as chemical ones. As a result,
considering these new forces, it is reasonable to expect that in the
real dead protoplasm the protein molecules stick together to balls
or superclusters, and we should consider this effect in derivation
of  \(\Delta\tilde{F}(V,T)\). The following section generalizes the
derivation of corrections to the free energy in case of possible
clustering, and it becomes clear that if \(V_{KH}\) denotes the
volume of such a cluster and \(Q_v\) is a heat generation of the Van
der Waals model, and \(Q_{kv}\) is a heat generation for the same
model (considering the possible clustering), then
\begin{eqnarray}
Q_{kv}=Q_v\frac{V_{KH}}{V_{1H}}.\label{HEAT}
\end{eqnarray}

As a result, we have
\begin{eqnarray}
\frac{Q_{kv}}{C_V T}=54 \frac{V_{KH}}{V_H}
\end{eqnarray}

If this new formula considering collective phenomena is correct,
then we shall find that the dead protoplasm should contain \(\approx
225\) superclusters or aggregates, the properties of which we cannot
characterize yet because the conclusion about the number of clusters
is a result of a rather common theoretical analysis. On the other
hand, the protein aggregation in case of the cell death is a
well-known phenomenon.

\section{Van der Waals Model of Protoplasm. Numerical Evaluations 2. Clustering of Protein Molecules}
In this section we estimate a correction to the free protein energy
\(\Delta F\) with respect to their clustering. The correction
\(\Delta F(V,T,N)\) is
\begin{eqnarray}
\Delta F=-T \ln \int \frac {d^3
q_1}{V}...\frac{d^3q_N}{V}e^{-\frac{U(q)}{T}}.
\end{eqnarray}

Here \(q_1\),...,\(q_N\) are Cartesian coordinates of all molecules
having numbers \(1,...,N\) and symbol \(q\) means coordinates of all
molecules: \(q:=(q_1,...,q_N)\). \(U(q)\) is given by
\begin{eqnarray}
U(q)=U_u(q)+U_s(q).
\end{eqnarray}
Here \(U_u\) is a potential energy of the Van der Waals attraction
between particles:
\begin{eqnarray}
U_u(q)=\sum \limits_{1\leq i< j\leq N} V_u(q_i-q_j),\nonumber\\
\end{eqnarray}
where
\begin{eqnarray}
V_u(q)=-C\frac{1}{|q|^6}.
\end{eqnarray}
Hereafter \(C\) value is considered as a small parameter used to
make any asymptotical expansions.
\begin{eqnarray}
U_s(q)=U^1_s(q)+U^2_s(q),
\end{eqnarray}
where \(U^1_s(q)\)  is a potential energy of repulsion between
protein molecules, arising due to their volume is low bounded. For
example:
\begin{eqnarray}
U^1_s(q)=\sum \limits_{1\leq i< j\leq N} V^1_s(q_i-q_j),\nonumber\\
V^1_s(q)=0,\; {\rm if \mit} |q|>2r_0,\nonumber\\
V^1_s(q)=+\infty\; {\rm if \mit} |q|\leq 2 r_0,
\end{eqnarray}
\(r_0\) is a radius of the protein molecule. \(U^2_s\)  is a
potential energy of Casimir forces, arising in very closed distances
between protein molecules due to zero-point oscillations of the
electromagnetic field in the gap between separate proteins. These
forces are short-ranged but they have a high degree of cooperativity
which contributes to clusters formation.

Potentials \(U_s\), \(U_s^1\), \(U_s^2\) we sometimes call
superpotential, and forces corresponding to them, superforces, as
\(U_s^1\), \(U_s^2\) are much greater than \(U_u(q)\) and the
effective consideration of \(U_s\) in the Gibbs exponent comes down
to the fact that the whole configuration space of molecules is
replaced by its part.

Let's represent \(\Delta F\) in the following form
\begin{eqnarray}
\Delta F=\Delta F_0+\Delta F_1+...,
\end{eqnarray}
where \(\Delta F_0\) is a zero-order value of vanishing on \(C\),
\(\Delta F_1\) is a value of the first order of vanishing on \(C\)
and so on.

Let's begin from the definition of \(\Delta F_0\), in other words,
put \(C=0\) and \(U(q)=U_s(q)\).

Due to presence in \(U_s\) all the molecules stick together to
clusters (balls \(B\)) of \(N_B\) molecules in each. Let \(N_k\) be
a number of such balls (clusters). Obviously \(N = N_BN_k\).

By definition all the clusters identical and there are
\begin{eqnarray}
\frac{1}{N_k!} \frac{N!}{(N_B!)^{N_k}}
\end{eqnarray}
of equivalent ways to distribute molecules by clusters. Let's fix
one of these ways which we will follow hereafter, where molecules
having coordinates \(q_1,...,q_{N_B}\) belong to the first cluster
\(B_1\), molecules having coordinates \(q_{N_B+1},....,q_{2N_B}\)
belong to the second cluster \(B_2\) and so on.

During \(F_0\) calculation we naturally neglected interaction
between clusters, so:
\begin{eqnarray}
\Delta F_0=-T \ln \frac{N!}{N_k!} [\frac{1}{N_B!} \int
\frac{d^3q_1}{V}....\frac{d^3 q_{N_B}}{V} e^{-\frac {U_B(q_1,...,
q_{N_B})}{T}}]^{N_k},
\end{eqnarray}
where by definition we set:
\begin{eqnarray}
U_B(q_1,...,q_{N_B})=\sum \limits_{1\leq i<j\leq N_B}
V_s^1(q_i-q_j)+U_s^2(q_1,...,q_{N_B}).
\end{eqnarray}
We proceed from the assumption that the cluster is a ball and
bonding forces between molecules in the cluster are so strong that
the cluster volume equals to a sum of volumes of molecules it
consists of (that why cluster density can exceed the water density).

Further, one of variables \(q_1\),....,\(q_{N_B}\) is a center of
inertia of cluster \(B\). The choice of such a variable can be
performed in \(N_B\) ways. Supposing the center of the cluster
inertia is \(q_1\) we find by integrating it:
\begin{eqnarray}
\Delta F_0=-T \ln \frac{N!}{N_k!} [\frac{1}{(N_B-1)!} \int
\frac{d^3q_2}{V}....\frac{d^3 q_{N_B}}{V} e^{-\frac {U_B(0,...,
q_{N_B})}{T}}]^{N_k},
\end{eqnarray}
And the center of cluster \(B\) is 0.

Then,we assume every molecule in the cluster moves in an effective
field \(W\) and \( U_s^2(0,..., q_{N_B})=W (N_B-1)\). Instead of
integral \(\int \limits_{B\times...\times B} d^3q_2...d^3q_{N_B}\)
the presence of multiplier \(e^{-\frac{U_s^1(q_1,...,q_{N_B})}{T}}\)
imposes the use of integral
\begin{eqnarray}
\int \limits_{B\times...\times B}' d^3q_2...d^3q_{N_B},
\end{eqnarray}
where prime points at the incompressibility of molecules. Using
\(N!\approx (\frac{N}{e})^N\) we receive:
\begin{eqnarray}
\Delta F_0=-T \ln \frac{1}{N_k!} [\frac{1}{(N_B-1)!}
(\frac{N}{e})^{N_B}\int \limits_{B\times...\times B}'
\frac{d^3q_2}{V}....\frac{d^3 q_{N_B}}{V} e^{\frac{-W
(N_B-1)}{T}}]^{N_k}.
\end{eqnarray}
But integral \(\int \limits_{B\times...\times B}' d^3q_2....d^3
q_{N_B}\) is easy and equals to
\begin{eqnarray}
\int \limits_{B\times...\times B}' d^3q_2....d^3
q_{N_B}=(N_B-1)!(\frac{V_B}{N_B})^{N_B-1},
\end{eqnarray}
where \(V_B\) is the cluster volume and \(V_b:=\frac{V_B}{N_B}\) is
the volume of one protein molecule. As a result, we obtain:
\begin{eqnarray}
\Delta F_0=-T N_k\ln N_B [(\frac{V_B}{V_k})^{N_B-1}
\frac{1}{e^{N_B-1}} e^{\frac{-W (N_B-1)}{T}}],
\end{eqnarray}
where \(V_k:=\frac{V}{N_k}\) is the volume which falls on one
cluster.

Now, let's derive the equilibrium condition for the cluster \(B\)
from which we essentially can define the volume of this cluster.
Let's consider one equilibrium cluster in the volume \(V_k\). Now we
are generally interesting in configuration part of the free energy
since it splits from the kinetic one.

Let's add one more protein molecule to the volume \(V_k\). Its free
energy is
\begin{eqnarray}
f_1=-T \ln \int \limits_{V_k} \frac{1}{V_k} d^3q=0.
\end{eqnarray}
If this protein falls to the cluster, then \(N_B\) is increased by a
unit, but since the cluster is equilibrium \(\Delta F_0\) should not
change. In other words, the following equality should take place:
\begin{eqnarray}
\frac{d \Delta F_0}{d N_B}=0. \label{EQUILIB1}
\end{eqnarray}
The equilibrium condition (\ref{EQUILIB1}) is actually a common
thermodynamic condition of equilibrium if we suppose a number of
protein molecules in a cell \(N\) to be a parameter which has an
influence on the system condition. But we can use this condition
only if \(N\) is not preserved, put it otherwise, is not a motion
integral. But the fact that \(N\) is not a motion integrals follows
from the fact that in the living cell the quantity of protein
molecules does not remain invariant because of permanently
proceeding processes of proteins synthesis and degradation.

The equation (\ref{EQUILIB1}) leads to the equality:
\begin{eqnarray}
\frac{1}{N_B}+\ln
(\frac{V_B}{V_k})+(N_B-1)\frac{V_b}{V_B}-1-\frac{W}{T}=0.
\end{eqnarray}
In other words:
\begin{eqnarray}
\frac{V_B}{V_k}=e^{\frac{W}{T}}.
\end{eqnarray}
Finally:
\begin{eqnarray}
\Delta F_0=-T N_k \ln N_B +T[N-N_k].
\end{eqnarray}
As the number of clusters is small (\( N_k\ll N\)), then in square
bracket of the second summand \(N_k\) can be neglected as compared
with \(N\). Since \(N_B\) conversely is very large, we can neglect a
logarithmic term as compared with \(TN\). As a result, we find:
\begin{eqnarray}
\Delta F_0=TN.
\end{eqnarray}
This term just comes down to the additive renormalization of the
entropy and therefore in can be omitted.

Now, let's calculate \(\Delta F_1(V,T,N)\). We have:
\begin{eqnarray}
\Delta F_1=\Delta F-\Delta F_0+O(C^2).
\end{eqnarray}

Let's recall, that \(C\) is considered as as mall parameter. It
won't be a great mistake if instead of \(\Delta F_0\) we take
\(\Delta F_0\) calculated by \(N - 2\) molecules in the same volume
\(V\) at the same temperature \(T\). We have:
\begin{eqnarray}
\Delta F_1=-T \ln \int \frac{d^3 q_1}{V}....\frac{d^3 q_N}{V}
e^{\frac{-U(q)+\Delta F_0}{T}}+O(C^2)\nonumber\\
=-T \ln \int \frac{d^3 q_1}{V}....\frac{d^3 q_N}{V} \prod
\limits_{1\leq i<j\leq N} [e^{-\frac{V_u(q_i-q_j)}{T}}-1+1]
e^{\frac{-U_s(q)+\Delta F_0}{T}}+O(C^2)\nonumber\\
=-T\ln \int \frac{d^3 q_1}{V}....\frac{d^3 q_N}{V}\{1+\sum
\limits_{1\leq i<j\leq N}
[e^{-\frac{V_u(q_i-q_j)}{T}}-1]\}e^{\frac{-U_s(q)+\Delta
F_0}{T}}+O(C^2)\nonumber\\
=-T\ln \{1+ \int \frac{d^3 q_1}{V}....\frac{d^3 q_N}{V}\{\sum
\limits_{1\leq i<j\leq N}
[e^{-\frac{V_u(q_i-q_j)}{T}}-1]e^{\frac{-U_s(q)+\Delta
F_0}{T}}\}+O(C^2).
\end{eqnarray}
Eventually, we obtain:
\begin{eqnarray}
\Delta F_1=-T\ln \{1+ \frac{N^2}{2}\int \frac{d^3
q_1}{V}....\frac{d^3
q_N}{V}[e^{-\frac{V_u(q_1-q_2)}{T}}-1]e^{\frac{-U_s(q)+\Delta
F_0}{T}}\}+O(C^2).
\end{eqnarray}
Since the number of clusters \(N_k\) is still sufficiently large, we
neglect cases when \(q_1\) and \(q_2\) are in the same cluster and
suppose them lying in different clusters \(B_1\) and \(B_2\). Values
\(q_1\) and \(q_2\) are simultaneous centers of the corresponding
clusters with probability \(\frac{1}{N_B^2}\). Taking this into
account:
\begin{eqnarray}
\Delta F_1=-T\ln \{1+ \frac{N^2}{2} N_B^2\int \frac{d^3
q_1}{V}....\frac{d^3 q_N}{V}[\langle
e^{-\frac{V_u(q'_1-q'_2)}{T}}\rangle-1]e^{\frac{-U_s(q)+\Delta
F_0}{T}}\}+O(C^2).
\end{eqnarray}
Here variables \(q_1\) and \(q_2\) are already centers of inertia of
clusters \(B_1\) and \(B_2\). Variables \(q'_1\), \(q'_2\) are
equiprobably and independently distributed along clusters \(B_1\)
and \(B_2\) and angle brackets mean the averaging over positions
\(q'_1\) and \(q'_2\).

Considering the meaning of \(\Delta F_0\), the last equation can be
rewritten in the following way:
\begin{eqnarray}
\Delta F_1=-T\ln \{1+ \frac{N^2}{2V} N_B^2\int \limits_{|q|>2r_B}
d^3
q [\langle e^{-\frac{V_u(q'_1-q'_2)}{T}}\rangle-1] \nonumber\\
\times[\frac{N^{N_B-1}}{(N_B-1)!} \int \limits_{B\times...\times B}
\frac{d^3 q_2}{V}....\frac{d^3
q_{N_B}}{V}e^{\frac{-U_s(0,...,q_2)}{T}}]^2\}+O(C^2)=\nonumber\\
\Delta F_1=-T\ln \{1+ \frac{N^2}{2V} N_B^2\int \limits_{|q|>2r_B}
d^3
q [\langle e^{-\frac{V_u(q'_1-q'_2)}{T}}\rangle-1] \nonumber\\
\times[(\frac{V_B}{V_k})^{N_B-1}e^{\frac{-W(N_B-1)}{T}}]^2\}+O(C^2),
\end{eqnarray}
where  \(r_B\) is a radius of a cluster \(V_B=\frac{4 \pi}{3}
r_B^3\). But taking into account the clusters equilibrium conditions
we derived, the expression inside the last square bracket equals to
1. Eventually:
\begin{eqnarray}
\Delta F_1=-T\ln \{1+\frac{N^2}{2V} N_B^2\int \limits_{|q|>2r_B} d^3
q [\langle
e^{-\frac{V_u(q'_1-q'_2)}{T}}\rangle-1]\}+O(C^2)\nonumber\\
=-T \frac{N^2}{2V} N_B^2\int \limits_{|q|>2r_B} d^3 q [\langle
e^{-\frac{V_u(q'_1-q'_2)}{T}}\rangle-1]+O(C^2).
\end{eqnarray}
This formula assumes that cluster \(B_1\) has a center of inertia in
zero, \(q\) is a coordinate of center of inertia of cluster \(B_2\),
\(q'_1\) and \(q'_2\) are variables which are uniformly and
independently distributed over cluster \(B_1\) and \(B_2\)
respectively.

This formula particularly proves the rule (\ref{HEAT}) quoted in the
end of the previous section.

Now, let's define a part which "dead" protein in the protoplasm
occupies in the whole cell volume. Suppose the cluster of dead
protoplasm proteins is situated in the volume and has an
\(V':=\frac{V}{N_k}\) equilibrium volume. If the cluster has the
equilibrium size, then inserting one protein molecule to it needs a
zero work \(A\), in other words, an average potential energy of the
protein molecule in the cluster equals to zero. But all the protein
molecules are situated in the neutral external field outside the
cluster and in the field \(W\) inside the cluster. Let's consider
the chosen protein \(\mathfrak{B}\), it has three degrees of freedom
described by Cartesian coordinates \(q^1\), \(q^2\), \(q^3\) and
radius \(q^4\) (the protein can be compressed). The only thing we
can do is to take a positively defined quadratic form of coordinates
\(q^1,...,q^4\) as a raw estimation for the potential energy of
protein inside the cluster \(V(q^1,..., q^4)\):
\begin{eqnarray}
V(q^1,...,q^4)=W+\sum \limits_{i,j=1,...,4} (q^i-q^i_0)B_{i,j}
(q^j-q^j_0),
\end{eqnarray}
where \(q^1_0,...,q^4_0\) are equilibrium coordinates of the protein
molecule. For kinetic energy of the chosen protein \(\mathfrak{B}\)
as a raw estimation too we take a positively defined form of
velocities \(\dot{q}^1,...,\dot{q}^4\):
\begin{eqnarray}
E_{kin}=\sum \limits_{i,j=1,...,4} \dot{q}^i A_{i,j} \dot{q}^j
\end{eqnarray}
Note, that protein molecule in the water does not behave as an
oscillator along \(q_4\) because a water is incompressible.

But in the case of classical mechanics the average potential energy
of one-dimensional oscillator equals to \(\frac{1}{2}T\). Therefore,
the condition \(A = 0\) leads to \(W = -2T\). This equilibrium
condition cluster lets us find:
\begin{eqnarray}
\frac{V_B}{V'}=e^{\frac{W}{T}}=e^{-2}\approx0.136,
\end{eqnarray}
which is in a good agreement with true value
\(\frac{V_B}{V'}=0.163\) (see section 4).

\section{Superfluid Bose Gas on Protein Configuration Space as Model of Living Cell Protoplasm}
In this section we discuss the second model of the living cell
protoplasm, the superfluid Bose gas on the protein configuration
space. The common description of this model is given in the
introduction section. While investigating this model we'll show that
when the Ling's protoplasm dies, the folding of protein molecules
occurs, how the Ling's theory postulates on the qualitative level
(Ling, 2001).

The protoplasm is considered at zero temperature. This allows us to
apply essential tools of theoretical physics but conclusions made
for such extreme conditions are useful for understanding the
properties of the real cell.

Let's proceed to conclude our "superfluid" model. Suppose there are
\(N\) protein molecules in the protoplasm, \(x_i\), \(i=1,...,N\)
are coordinates of their centers of inertia and \(\sigma_i\) are
parameters, passing some space \(X\) with measured \(d\mu\)
describing inner freedom degrees of the protein molecule (its
configuration). Then, the state of \(N\) protein molecules is
described by the wave function
\(\Psi(x_1,\sigma_1,...,x_N,\sigma_N)\) depending on coordinates
\(x_i,\sigma_i\) of protein molecules. This wave function fulfills
the normalizing condition:
\begin{eqnarray}
\int
|\Psi(x_1,\sigma_1,...,x_N,\sigma_N)|^2dx_1,...,dx_nd\mu(\sigma_1)...d\mu(\sigma_N)=1
\end{eqnarray}
and symmetry condition \(\forall P \in S_N\) permutation of the set
of \(N\) elements
\begin{eqnarray}
(\hat{P}\Psi)(x_1,\sigma_1,...,x_N,\sigma_N):=\Psi(x_{P(1)},\sigma_{P(1)},...,x_{P(N)},\sigma_{P(N)})=\chi(P)
\Psi(x_1,\sigma_1,...,x_N,\sigma_N),
\end{eqnarray}
where  \(\chi(P)\equiv 1\) if protein molecules satisfy to the Bose
--- Einstein statistics, and \(\chi(P)=\rm sgn \mit (P)\) (signum of
permutation) if protein molecules satisfy to the the Fermi
--- Dirac statistics.

To describe this system, by using common considerations we'll try to
find its Hamiltonian as per the adiabatic limit method (see the
previous section). Suppose \(V(x_1,\sigma_1,...,x_N,\sigma_N)\) is a
change of the smallest eigenvalue of the protoplasm Hamiltonian
while adding there \(N\) protein molecules having coordinates
\(x_1,\sigma_1,...,x_N,\sigma_N\), and \(\hat{V}\) is an operator of
multiplication of wave function of our system on
\(V(x_1,\sigma_1,...,x_N,\sigma_N)\). Then, obviously, the effective
Hamiltonian of our system is
\begin{eqnarray}
\hat{H}_{eff}=\sum \limits_{i=1}^N \hat{T}_i+\hat{V},
\end{eqnarray}
where \(\hat{T}_i\) is a kinetic energy operator of \(i\)-th protein
as a material point:
\begin{eqnarray}
\hat{T}_i=-\frac{1}{2M}\nabla_i^2.
\end{eqnarray}
To simplify our model let's suppose \(\hat{V}\) describes only a
pair interaction of protein molecules.
\begin{eqnarray}
\hat{V}=\sum \limits_{i=1}^N \hat{V}_i+\sum \limits_{N \geq i>j\geq
0} \hat{V}_{i,j},
\end{eqnarray}
where in Dirac notation
\begin{eqnarray}
\langle
\sigma_1,x_1,...,x_n,\sigma_n|\hat{V}_i|x_1,\sigma'_1,...,x_n\sigma'_n\rangle=U_1(x_i,\sigma_i|x'_i,\sigma'_i)
\end{eqnarray}
for some function \(U_1(x'_i,\sigma'_i|x'_i,\sigma'_i)\) fulfilling
obvious conditions arising from the Hermitian nature of
\(\hat{V}_i\). Similarly
\begin{eqnarray}
\langle\sigma_1,x_1,...,x_n,\sigma_n|\hat{V}_{i,j}|x_1,\sigma'_1,...,x_n\sigma'_n\rangle=U_2(x_i,\sigma_i,x_j,\sigma_j|x'_i,\sigma'_i,x'_j,\sigma'_j)
\end{eqnarray}
for some function
\(U_2(x_i,\sigma_i,x_j,\sigma_j|x'_i,\sigma'_i,x'_j,\sigma'_j)\)
fulfilling evident conditions arising from the Hermitian nature of
\(\hat{V}_{i,j}\).

We consider the protoplasm at the zero temperature (measured in the
energy scale). The following objection may arise against this
acceptance: at zero temperature all the metabolic process in cell
are stopped and the cell dies (it is the way how the non-equilibrium
thermodynamics understands the life and the death). However, it is
well-known that cells (even embryos including human ones) come back
to life after freezing in the liquid nitrogen when there are no
flows of matter or energy at all. Our approach differs: we consider
the living state as stationary and thermodynamically sustainable,
but not the equilibrium one. Precisely this state is "freezing" and
can keep its properties even at the absolute zero. That's why the
frozen cell comes back to life when temperature conditions became
normal again. The thermodynamic identity of the resting state of the
living and cells allows us to apply the same analytical apparatus.

As the average kinetic energy of a heat motion of the protein
molecule \(T_i\) at temperature \(T\) is equal to \(kT\), so at the
zero temperature kinetic energy equals to zero. Therefore, we can
omit the term corresponding to kinetic energy from the Hamiltonian
\(H_{eff}\). In other words, when \(T =0\) we have the following
expression for \(H_{eff}\):
\begin{eqnarray}
\hat{H}_{eff}=\sum \limits_{i=1}^N \hat{V}_i+\sum \limits_{N \geq
i>j\geq 0} \hat{V}_{i,j}.
\end{eqnarray}

With regard to the protein molecules themselves, according to Ling's
model of the cell (Ling, 2001), they are situated in points of a
lattice and emerged into the volume of water associated with these
proteins (the water is as well structured and ordered), while
\(i\)-th point has coordinates \(x_i\).
\(\Psi(x_1,\sigma_1,...,x_N,\sigma_N)\) has the following form:
\begin{eqnarray}
\Psi(x_1,\sigma_1,...,x_N,\sigma_N)=C\sum \limits_{P \in S_N} \prod
\limits_{i=1}^N\delta(x'_i-x_{P(i)})A^P(\sigma_1,...,\sigma_N)\chi(P),
\label{Anzatz}
\end{eqnarray}
where \(A^P\) fulfills the following property arising from the
property of \(\Psi(x_1,\sigma_1,...,x_N,\sigma_N)\) to fulfill the
Bose-Einstein or Fermi-Dirac statistics:
\begin{eqnarray}
A^{PQ^{-1}}(\sigma_{Q(1)},....,\sigma_{Q(N)})=A^P(\sigma_1,...,\sigma_N),
\end{eqnarray}
and a divergent constant \(C\) is chosen from the condition
\begin{eqnarray}
C \int dx_1,...,dx_N(\prod \limits_{i=1}^N\delta(x_i-x'_i))^2=1.
\end{eqnarray}

At \(T = 0\) the whole protein system is in the ground state
\(\Psi_0(x_1,\sigma_1,...,x_N,\sigma_N)\) which can be defined based
upon the requirement of minimality of the averaged Hamiltonian
\(H_{eff}\) over all normalized states. In other words, in the state
\(\Psi_0\) the averaged value \(\langle
\Psi|\hat{H}_{eff}|\Psi\rangle\) of the effective Hamiltonian
reaches its minimum value
\begin{eqnarray}
E_0=\min \limits_{\|\Psi\|=1}\langle \Psi|\hat{H}_{eff}|\Psi\rangle.
\end{eqnarray}
Taking into account the aforesaid, in searching the minimum value of
\(\langle \Psi|\hat{H}_{eff}|\Psi\rangle\) we can restrict ourselves
to states of the form (\ref{Anzatz}). Since for states of the form
(\ref{Anzatz}) the coordinates of inertia centers of proteins are
localized, we may say that \(\hat{V}_i\) acts only in \(L^2(X,d
\mu)\) and \(\hat{V}_{i,j}\) acts in \(L^2(X^2,d \mu\times d\mu)\).
Then, our variational problem comes down to finding the functional
minimum
\begin{eqnarray}
\int
d\mu(\sigma_1)...d\mu(\sigma_n)A^{\star}(\sigma_1,...,\sigma_N)\{(\sum
\limits_{i=1}^N \hat{V}_i+\sum \limits_{N \geq i>j\geq 0}
\hat{V}_{i,j})A\}(\sigma_1,...,\sigma_N)
\end{eqnarray}
upon the additional condition
\begin{eqnarray}
\int|A(\sigma_1,...,\sigma_n)|^2d\mu(\sigma_1)...d\mu(\sigma_N)=1.
\end{eqnarray}

Let's introduce a function
\(V_2(\sigma_1,\sigma_2|\sigma'_1,\sigma'_2|x'_i,x'_j)\)  with a
formula
\begin{eqnarray}
 V_2(\sigma_1,\sigma_2|\sigma'_1,\sigma'_2|x'_i,x'_j)=U_2(x'_i,\sigma_1,x'_j,\sigma_2|x'_i,\sigma'_1,x'_j,\sigma'_2).
\end{eqnarray}
Then, in obvious notations
\begin{eqnarray}
 \langle
 \sigma_1,...,\sigma_N|\hat{V}_{i,j}|\sigma'_1,....,\sigma'_N\rangle=V_2(\sigma_i,\sigma_j|\sigma'_i,\sigma'_j|x_i,x_j).
\end{eqnarray}
Let \(P\) be any permutation from \(S_N\). Let's define the operator
\(\hat{V}_{i,j}^P\) with the following formula
 \begin{eqnarray}
 \langle
 \sigma_1,...,\sigma_N|\hat{V}_{i,j}^P|\sigma'_1,...,\sigma'_N\rangle=V_2(\sigma_i,\sigma_j|\sigma'_i,\sigma'_j|x_{P(i)},x_{P(j)}).
 \end{eqnarray}
Let by definition
\begin{eqnarray}
 \hat{V}^e_{ij}:=\frac{1}{|S_N|} \sum \limits_{P \in S_N}
 \hat{V}_{i,j}^P.
\end{eqnarray}

Let's renumber protein molecules in such a way that the protein
which starts the numeration is situated sufficiently far from the
cell surface. Then

\begin{eqnarray}
\langle\sigma_1,...,\sigma_n|\hat{V}_{ij}^e|\sigma'_1,...,\sigma'_n\rangle=\frac{1}{N}\sum
\limits_{k=1}^N V_2(\sigma_i,\sigma_j|\sigma'_i,\sigma'_j|x_0,x_k);
\end{eqnarray}

Let by definition

\begin{eqnarray}
U_2(\sigma_1,\sigma_2|\sigma'_1,\sigma'_2):=\frac{1}{N}\sum
\limits_{k=1}^N V_2(\sigma_1,\sigma_2|\sigma'_1,\sigma'_2|x_0,x_k).
\end{eqnarray}

Now, we'll show that within the framework of some approximation the
class of functions, where solution of variational problem we just
set up is searched, can be narrowed down to symmetrical functions,
i.e. such functions that \(\forall P \in S_N\)
\begin{eqnarray}
A^P(\sigma_1,...,\sigma_N)=A^{id}(\sigma_1,...,\sigma_N)=:A(\sigma_1,...,\sigma_N).
\end{eqnarray}
We suppose that \(\hat{V}_{i,j}\) is proportional to real constant
\(\lambda\) which is a small parameter. Therefore, if parameter
\(\lambda\) is enough small, the different protein molecules can be
considered as non-correlated ones, in other words, behaving
independently. This means that the wave function
\(A(\sigma_1,...,\sigma_N)\) can be represented in the following
form:
\begin{eqnarray}
A(\sigma_1,...,\sigma_N)=\prod \limits_{i=1}^N \xi(\sigma_i)
\end{eqnarray}
for a function \(\xi(\sigma)\in L_2(X,d\mu)\) fulfilling the
following normalizing condition:
\begin{eqnarray}
\int |\xi(\sigma)|^2d\mu(\sigma)=1.
\end{eqnarray}

Let \(\hat{U}^1\) be an operator in \(L_2(X,d\mu)\) which have
matrix elements in \(\sigma\)-representation defined by function
\(U_1\), and \(\hat{U}^2\) be an operator in \(L_2(X\times
X,d\mu\times d\mu)\) which have matrix elements in
\(\sigma\)-representation defined by function \(U_2\). Let's denote
\(U_2(\xi)\) as an operator in \(L_2(X,d\mu)\) defined from the
following relation:
\begin{eqnarray}
\langle f|U_2(\xi)|g\rangle=\langle f\otimes
\xi|U_2|g\otimes\xi\rangle.
\end{eqnarray}
The approximation we've just described is called a mean field
approximation and time evolution of function \(\xi\) in it is
defined by the following equation:
\begin{eqnarray}
i\frac{\partial}{\partial t}\xi=\hat{V}^1\xi+N\hat{V}^2(\xi)\xi.
\end{eqnarray}
Any how, in our approximation
\begin{eqnarray}
A(\sigma_1,...,\sigma_N)\sim\prod \limits_{i=1}^N \xi(\sigma_i).
\end{eqnarray}

As the class of test functions is narrowed down to the symmetrical
ones in our variational problem, then, instead of \(\hat{H}_{eff}\)
minimum eigenvalue calculation, we can seek the minimum eigenvalue
of the Hamiltonian:
\begin{eqnarray}
\hat{H'}_{eff}=\sum \limits_{i=1}^N \hat{V}_i+\sum \limits_{N \geq
i>j\geq 0} \hat{V}_{i,j}^e.
\end{eqnarray}

Within the framework of our approximation the energy of the ground
state of \(\hat{H}'_{eff}\) is equal to the energy of the ground
state of \(\hat{H}_{eff}\).

\textbf{Proposition.} There exist functions
\(U'_1(\sigma_1|\sigma'_1)\),
\(U'_2(\sigma_1,\sigma_2|\sigma'_1,\sigma'_2)\) such as if
\(\hat{V'}_i\), \(\hat{V}'_{i,j}\) operators defined by the
following relations
\begin{eqnarray}
\langle
\sigma_1,....,\sigma_N|\hat{V}'_{i}|\sigma'_1,...,\sigma'_N\rangle=U'_1(\sigma_i|\sigma'_i),\nonumber\\
\langle\sigma_1,....,\sigma_N|\hat{V}'_{ij}|\sigma'_1,...,\sigma'_N\rangle=U'_2(\sigma_i,\sigma_j|\sigma'_i,\sigma'_j),
\end{eqnarray}
the operators \(V'_i\) and \(V'_{ij}\) are Hermitian ones,
\begin{eqnarray}
\hat{H'}_{eff}=\sum \limits_{i=1}^N \hat{V}'_i+\sum \limits_{N \geq
i>j\geq 0} \hat{V'}_{i,j}.
\end{eqnarray}
and operators \(V'_i\) and \(V'_{ij}\) fulfill one more condition
being as follows. Let \(\{\varphi'_n|n=0,1,...\}\)be an orthonormal
basis of eigenfunctions of operator  \(\hat{U}'_1\) in
\(L_2(X,d\mu)\) defined by the following equation:
\begin{eqnarray}
(\hat{U}'_1\varphi)(\sigma)=\int
U'_1(\sigma,\sigma')\varphi(\sigma')d\mu(\sigma').
\end{eqnarray}
Let \(U'_2\) is the operator in \(L_2(X\times X,d\mu \times d\mu)\)
specified by the relation
\begin{eqnarray}
(\hat{U}'_2\varphi)(\sigma_1,\sigma_2)=\int
U'_2(\sigma_1,\sigma_2|\sigma'_1,\sigma'_2)\varphi(\sigma'_1,\sigma'_2)d\mu(\sigma)d\mu(\sigma').
\end{eqnarray}
Let \((\hat{U'}_2)_{mn,m'n'}\) be a matrix element of operator \(\hat{U'}_2\) with
respect to the basis \(\{\varphi_i\otimes\varphi_j|i,j=0,1,...\}\).
\begin{eqnarray}
(\hat{U}'_2)_{mn,m'n'}=\langle
\varphi_m\otimes\varphi_n|\hat{U}'_2|\varphi_{m'}\otimes\varphi_{n'}\rangle.
\end{eqnarray}
Then, for any \(n = 1,2,3...\)
\begin{eqnarray}
(U'_2)_{00,0n}=0
\end{eqnarray}
and similar equalities take place where index \(n\) stands on
remaining three places and other have zeros.

\textbf{Proof.} Let \(\hat{U}_1\) be an operator in specified by the
following relation:
\begin{eqnarray}
(\hat{U}'_1\varphi)(\sigma)=\int
U'_1(\sigma,\sigma')\varphi(\sigma')d\mu(\sigma'),
\end{eqnarray}
and let \(\{\varphi_n|n=0,1,...\}\) be an orthonormal basis of
eigenfunctions of operator \(\hat{U}_1\).

We represent the operator \(\hat{V}^e_{i,j}\) as a sum of three
summands:
\begin{eqnarray}
\hat{V}^e_{i,j}=\hat{V}^{e,1}_{i,j}+\hat{V}^{e,2}_{i,j}+\hat{V}^{e,3}_{i,j},
\end{eqnarray}
where operators \(\hat{V}^{e,1}_{i,j}\), \(\hat{V}^{e,2}_{i,j}\),
\(\hat{V}^{e,3}_{i,j}\) are based on functions
\((U^1_2)(\sigma_1,\sigma_2)\), \((U^2_2)(\sigma_1,\sigma_2)\),
\((U^3_2)(\sigma_1,\sigma_2)\) defined below by the principle which
the \(\hat{V}^e_{ij}\) is based on
\(U_2(\sigma_1,\sigma_2|\sigma'_1,\sigma'_2)\). The functions
\(U^1_2,\;U^2_2,\;U^3_2\) are defined by the following condition.
Let \(\hat{U}^1_2,\;\hat{U}^2_2,\;\hat{U}^3_2\) be operators based
on functions \(U^1_2,\;U^2_2,\;U^3_2\) using the same scheme as used
by the operator \(\hat{U}_2\) based on the function \(U_2\). Among
all matrix elements of operators \(\hat{U}^1_2\), \(\hat{U}^2_2\)
only the following elements are non-zero: \((U_2^1)_{00,0m}\),
\((U_2^1)_{00,m0}\), \((U_2^2)_{m0,00}\), \((U_2^2)_{0m,00}\),
\(m=1,2,...\) and among all matrix elements of the operator
\(\hat{U}_3^2\) only remained elements are non-zero. It is easily
seen that if we redefine \(\hat{V}^{e,3}_i\) in a proper manner,
then we can include summands corresponding to \(\hat{V}^{e,1}_{i,j}\),
\(\hat{V}^{e,2}_{i,j}\) into \(\hat{V}_{i}\).

But \(\hat{V}_{ij}\) and, consequently, \(\hat{V}_{ij}^e\) are
first-order values by the constant of protein interaction
\(\lambda\). Therefore, in just described substitution, the basis of
eigenfunctions of operator \(\hat{U}_1\)
\(\{\varphi_n|n=0,1,2,...\}\) passes to the basis of eigenfunctions
of redefined operator \(\hat{U}'_1\) \(\{\psi_n|n=0,1,2,...\}\) such
that \(\varphi_n-\psi_n\) are the first-order of vanishing values by
a coupling constant \(\lambda\) \(\forall n=0,1,2,...\).

The matrix elements of just redefined \(\hat{V}^e_{ij}\) of the form
\((\hat{U}_2)_{00,0m}\) \(m=1,2,...\) (\(\hat{U}_2\) is coupled with
\(\hat{V}^e_{ij}\) by the relation we've mentioned above) relating
to the basis \(\{\varphi_n\}\) are equal to zero. But since
\(\hat{V}^e_{ij}\) are of first-order of vanishing by the
interaction constant, then matrix elements of the form
\((\hat{U}_2)_{00,0m}\) of redefined \((\hat{U}_2)\) related
to the basis \(\{\psi_n\}\) are of the second order
of vanishing by the coupling constant.

The procedure described above can be continued endlessly by
sequentially lowering the order of vanishing of those terms we want
to omit in the Hamiltonian.

The proposition may be considered as a proven one even without
discussing a problem of convergence of the above described iteration
procedure, as all formulas which may be derived hereafter will be
just asymptotic expansions in a small parameter \(\lambda\).

So, the effective Hamiltonian for considered system of protein
molecules, which we want to use for defining the ground state of our
system, has the following form:
\begin{eqnarray}
H'_{eff}=\sum \limits_{i=1}^N \hat{V}_i+\sum \limits_{1\leq i\leq
j\leq N} \hat{V}^e_{ij}. \label{effHam}
\end{eqnarray}

Here operators \(\hat{V}_i\) have "act" only on coordinate \(\sigma_i\) of the
wave function and differ from each other only by a number of the
coordinate on which they "act". Operators \(\hat{V}_{ij}\) "acts" only on coordinates
of numbers \(i\) and \(j\) of the wave function and differ from each
other only by numbers of coordinates on which they "acts".

Operator \(\hat{V}_{ij}\) "acts" only on coordinates of numbers
\(i\) and \(j\) of the wave function and differ from each other only
by numbers of coordinates on which they "acts".

If in (\ref{effHam}) we evidently extract the dependence from the
interaction constant and number of particles, then we shall find
\begin{eqnarray}
H_{eff}=\sum \limits_{i=1}^N \hat{V}_i+\frac{\lambda}{N}\sum
\limits_{1\leq i\leq j\leq N} \tilde{V}^e_{ij},
\end{eqnarray}
where prime of \(H_{eff}\) is omitted and \(\tilde{V}^e_{ij}\) does
not depend on \(N\) (asymptotically at \(N\rightarrow \infty\)).

We intend to investigate the Hamiltonian \(H_{eff}\) using the
apparatus of secondary quantization.

Let \(\{\varphi_n|n=0,1,2...\}\) be still an orthonormal basis of
eigenfunctions of operator \(\hat{U}_1\). Let
\(\mathcal{F}=\Gamma(L_2(X,d\mu))\) be a boson Fock space over
\(L_2(X,d\mu)\). Let \(a^+_i,\;a_i\) be operators of particles
creation and annihilation in the state \(\varphi_i\) acting in
\(\mathcal{F}\). \(\forall i=0,1,...\) operators \(a^+_i\) and
\(a_i\) are conjugated to each other and fulfill the following
canonical commutating relations:
\begin{eqnarray}
{[a_i,a_j]=[a^+_i,a^+_j]=0},\;i,j=0,1,2...,\nonumber\\
{[a_i,a^+_j]}=\delta_{ij},
\end{eqnarray}
where \(\delta_{ij}\) is a common Kronecker delta. We suppose that
eigenfunctions \(\varphi_n\) are numerated in such a way that
\(E_n\) increases with increasing of number \(n\). In representation
of the secondary quantization the Hamiltonian \(H_{eff}\) is given
by:
\begin{eqnarray}
H_{eff}=\sum \limits_{n=0}^{\infty} E_n
a^+_na_n+\frac{\lambda}{2N}\sum \limits_{m,n,m',n'}a^+_ma^+_n
(U_2)_{mn,m'n'}a_{m'}a_{n'}.\label{effHam1}
\end{eqnarray}
But it is a common Hamiltonian studied in the superfluid theory, and
we can use the same methods for its investigation.

At \(T = 0\) a portion of proteins is in the ground state
\(\varphi_0\). Let \(N_0\) be a number of protein molecules in the
ground state. As \(\frac{\lambda}{N}\ll1\), then
\(\frac{N-N_0}{N_0}\ll1\). Because of this reason in the right part
of relation \(a_0a^+_0-a^+_0a_0=1\) we can neglect a unity and
suppose \(a_0\) and \(a^+_0\) are usual \(c\)-numbers.

So, we can simply suppose that \(a_0=\sqrt{N}e^{-i\varphi}\) and
\(a^+_0=\sqrt{N}e^{i\varphi}\)  and for some real number
\(\varphi\).

Now let's consider the second summand in the right part of
(\ref{effHam1}). According to the assumption proven above we can
suppose that this summand has no terms linear by \(a_i^+,\;a_i\),
\(i=1,2,...\). Quadratic terms have an order of \(\lambda\), cubic
\(\frac{\lambda}{\sqrt{N}}\), and quartic \(\frac{\lambda}{N}\).
Therefore, supposing the number of proteins in the cell is finite
but a sufficiently large and interaction constant \(\lambda\) is
small, in the Hamiltonian (\ref{effHam1}) we can keep only quadratic
terms on operators \(a_i^+,\;a_i\), \(i=1,2,...\).

Consequently, the effective Hamiltonian \(H_{eff}\) is given by:
\begin{eqnarray}
H_{eff}=\pi c N_0+\sum
\limits_{n=1}^{\infty}E_na^+_na_n\nonumber\\
+{\lambda}\{\sum
\limits_{m,n=1}^{\infty}A_{mn}e^{-2i\varphi}a^+_ma^+_n+\sum
\limits_{m,n=1}^{\infty}A_{mn}^{\star}e^{2i\varphi}a_m a_n+\sum
\limits_{m,n=1}^{\infty}B_{mn}a^+_ma_n\}. \label{HamSupp}
\end{eqnarray}
Here \(c\) is a material constant (does not depend on \(N_0\)),
\(A_{mn}\), \(B_{mn}\) are some matrixes \(A_{mn}=A_{nm}\),
\(B^{\star}_{mn}=B_{nm}\), \(\star\) is a sign of complex
conjugation.

There is a classical variable \(\varphi\) which is canonically
conjugated to variable \(J=\pi N_0\). Variable \(\varphi\) satisfies
the equation \(\dot{\varphi}=c\) and, therefore, value
\(\psi=\varphi-ct\) (\(t\) is time) is a motion integral. Whole our
system (denoted by \(M\)) has a structure of a direct product of two
systems \(M=M_1\times M_2\) where, roughly speaking, the system
\(M_1\) is described by variables \(\varphi\) and \(J\), and the
\(M_2\) system is described by (noncommutative) variables
\(a^+_n,\;a_n\), \(n=1,2...\).

Now, let's consider two cases: when the cell is a live and when it
is dead. In the case of the living cell, the expression under the
integral for statistical weight contains a multiplier
\(\delta(\varphi-\varphi')\), \(\varphi'\) is a fixed real number,
and now the free energy \(F(\varphi')\) is a function of
\(\varphi'\). However, one can readily see that the free energy
\(F(\varphi')\) does not depend on \(\varphi'\). Indeed, the change
of \(\varphi'\) to the value \(\delta\varphi\) can really be
compensated by proper canonical transformation of operators
\(a^+_n,\;a_n\), \(n=1,2...\):
\begin{eqnarray}
a^+_n\mapsto a^+_ne^{i\delta\varphi}, \nonumber\\
a_n\mapsto a_ne^{-i\delta\varphi}.
\end{eqnarray}

Now, integral \(\varphi\) fulfills the equivalence principle we have
considered above (Prokhorenko and Matveev, 2011). As the formula of
the generalized microcanonical distribution includes a multiplier
factor  \(\delta(\varphi-\varphi')\), the \(M_2\) system may be
described by using the Hamiltonian (\ref{HamSupp}) where \(\varphi\)
is given as a some value and \(N_0\) is given taking into account
that the total system particles number is equal to \(N\).

Now, let's consider a dead cell. In this case the cell is described
by means of the equilibrium Gibbs distribution. Let's write out
Hamilton's equations for \(\varphi\) and \(J\). So, we have:
\begin{eqnarray}
\dot{\varphi}=c,\nonumber\\
\dot{J}=0+\lambda L(a^+,a),
\end{eqnarray}
where \(L(a^+,a)\) is a quadratic function of operators
\(a^+_n,\;a_n\), \(n=1,2...\). But if \(N\rightarrow \infty\), then
\(J\sim N\), and \(L(a^+,a)\sim 1\). Therefore, when \(N\) values
are high enough, \(L(a^+,a)\) value may be neglected in the
Hamiltonian equations for \(\varphi\)  and \(J\) and the following
equations are obtained:
\begin{eqnarray}
\dot{\varphi}=c,\nonumber\\
\dot{J}=0.
\end{eqnarray}

Thus, when \(N\) values are high enough, dynamics of the \(M_1\)
system separates from the dynamics of \(M_2\). This means that dynamical
variables for \(M_1\) and \(M_2\) systems are independent and the
probability distribution for the \(M_1\) system, corresponding to
Gibbs distribution for the whole system, is defined by the formula:
\begin{eqnarray}
\rho(\varphi,J)=\rm const \mit \delta(J-J').
\end{eqnarray}

Now, let's define the distribution function for the system \(M_2\).
For this we use a classical analogy. Suppose the Hamiltonian system
\(K\) has a structure of direct product of systems \(K_1\) and
\(K_2\), \(K = K_1 \times K_2\) and \(H(x,y)\) is the Hamiltonian of
the whole system, \(x\in K_1,\;y\in K_2\). If the \(K\) system is
described by the Gibbs canonical distribution corresponding to \(T\)
temperature, then the probability distribution for the \(K_2\)
system is the following:
\begin{eqnarray}
\rho(y)=\frac{1}{Z}e^{-\frac{F(y|T)}{T}},
\end{eqnarray}
where
\begin{eqnarray}
F(y|T)=-T\ln \int d\Gamma_x e^{-\frac{H(x,y)}{T}},
\end{eqnarray}
and \(Z\) is a normalized factor. Alternately, as can be seen above,
the effective Hamiltonian for the \(K_2\) system can be obtained in
the following way. Let's write out the Hamiltonian equations for the
system \(K_2\). Assume \(p\) and \(q\) are canonical coordinates and
momenta for the system \(K_2\) system. Then:
\begin{eqnarray}
\dot{p}=-\frac{\partial H(x,y)}{\partial q}, \nonumber\\
\dot{q}=\frac{\partial H(x,y)}{\partial p}.
\end{eqnarray}

Now, if we average right parts of the two last equations by the
conditional Gibbs distribution for \(K_1\) \(w(x|y)\) having a
specified value  \(y \in K_2\), then we obtain closed with respect
to \(y\) Hamiltonian equations where \(F(y|T)\) is the Hamiltonian.

Therefore, in order to define the effective dynamics for the \(M_2\)
system for a dead cell, the following method, reasoning by analogy,
can be used. Heisenberg equations for \(a^+_n,\;a_n\), \(n=1,2...\)
can be written out and averaged over \(\varphi\)  and \(J\).
However, as shown above, if  \(N\rightarrow\infty\), then
\(\varphi\) is independent of \(a^+_n,\;a_n\), \(n=1,2...\) and
distribution for \(\varphi\) is uniform. Direct calculation shows
that if the right parts of Heisenberg equations for \(a^+_n,\;a_n\),
\(n=1,2...\) are averaged by the uniformly distributed  \(\varphi\),
we obtain Heisenberg equations where the Hamiltonian is:
\begin{eqnarray}
H_{eff}=cN_0+\sum
\limits_{n=1}^{\infty}E_na^+_na_n\nonumber\\
+{\lambda}\sum \limits_{m,n=1}^{\infty}B_{mn}a^+_ma_n.
\end{eqnarray}

It's evident, when \(T = 0\) \(N = N_0\), i.e. all protein molecules
are folded.

Now, let's pass to analysis of \(H_{eff}\) using a well-known theory
of normal forms of quadratic Hamiltonians. For Hamiltonians of a
general type this theory was developed by H. Poincare and G.D.
Birkhoff (see Arnold, 2003) and for quantum mechanics it was adapted by
N.N. Bogoliubov (Bogoliubov and Bogoliubov (jr), 1984). This theory
was established only for dynamical systems with a finite number of
degrees of freedom, but we use it for systems with an infinite
number of degrees of freedom, as it is a standard practice in
physics. For this purpose we replace the Hamiltonian \(H_{eff}\) by
cuted Hamiltonian \(H^c_{eff}\)  describing the system having \(L\)
degrees of freedom:
\begin{eqnarray}
H_{eff}^c=\pi cN_0+\sum
\limits_{n=1}^{L}E_na^+_na_n\nonumber\\
+{\lambda}\{\sum
\limits_{m,n=1}^{L}A_{mn}e^{-2i\varphi}a^+_ma^+_n+\sum
\limits_{m,n=1}^{L}A_{mn}^{\star}e^{2i\varphi}a_m a_n+\sum
\limits_{m,n=1}^{L}B_{mn}a^+_ma_n\}. \label{HamSup}
\end{eqnarray}
In this case, the theory of quadratic Hamiltonians confirms that for
general matrixes \(A_{mn}\), \(B_{mn}\) there is such a linear
transformation of creation and annihilation operators
\(a^+_n,\;a_n\), \(n=1,2...\) to operators \(\xi^+_n,\;\xi_n\),
\begin{eqnarray}
\xi^+_n=\sum \limits_{m=1}^L U_{nm}a^+_m+\sum \limits_{m=1}^L
V_{nm}a_m,\nonumber\\
\xi_n=\sum \limits_{m=1}^L U^{\star}_{nm}a_m+\sum
\limits_{m=1}^L V^{\star}_{nm}a^+_m\nonumber\\
\end{eqnarray}
that \(\forall n=1,2...\) \(\xi_n\) is conjugated to \(\xi_n^+\);
this transformation is canonical:
\begin{eqnarray}
{[\xi_n,\xi_m]}={[\xi^+_n,\xi^+_m]}=0,\nonumber\\
{[\xi_n,\xi_m^+]}=\delta_{nm},\;m,n=1,2,3...
\end{eqnarray}
and using new variables, Hamiltonian \(H^c_{eff}\) is:
\begin{eqnarray}
H^c_{eff}=\sum \limits_{i=1}^K\omega_n\xi^+_n\xi_n+\sum
\limits_{i=K+1}^L\chi_n(\xi^+_n\xi^+_n+\xi_n\xi_n).\label{CEH}
\end{eqnarray}
 \(\omega_n\), \(\chi_n\) are some real numbers. However, the physical considerations
 make clear that \(H^c_{eff}\) should be low bounded. Let's denote \(\mathcal{H}^c\) as
 Hilbertian space where Hamiltonian \(H^c_{eff}\) acts. This Hilbertian space
 is represented by the tensor product of Hilbertian spaces
 \(\mathcal{H}^c_i\)
 \begin{eqnarray}
\mathcal{H}^c=\bigotimes \limits_{i=1}^L
\mathcal{H}^c_i,\label{TENSOR}
\end{eqnarray}
where every \(\mathcal{H}^c_i\) is isomorphic to \(L_2(R)\) and this
isomorphism can be chosen in such a way that the following condition
is fulfilled. In \(L_2(R)\) the operators of coordinate \(\hat{q}\)
and momenta \(\hat{p}\) operate in the standard way:
\begin{eqnarray}
\hat{p}\hat{q}-\hat{q}\hat{p}=\frac{1}{i}.
\end{eqnarray}
\(\xi^+_n\) and \(\xi_n\) operate only on the \(n\)-th multiplier in
(\ref{TENSOR}). The isomorphism between \(\mathcal{H}^c_i\) and
\(L_2(R)\) mentioned above can be chosen in such a way that due to
this is isomorphism the following is correct:
\begin{eqnarray}
\xi_n=\frac{1}{\sqrt{2}}(\hat{p}-i\hat{q}),\nonumber\\
\xi_n^+=\frac{1}{\sqrt{2}}(\hat{p}+i\hat{q}).
\end{eqnarray}
But due to the isomorphism between \(\mathcal{H}^c_n\) and
\(L_n(R)\) mentioned above the operator
\(\xi^+_n\xi^+_n+\xi_n\xi_n\) equals to the operator
\(\hat{p}^2-\hat{q}^2\). Evidently, the last operator is not neither
low nor upper bounded. Therefore, the physical requirement of
positiveness of \(H^c_{eff}\) results in the fact that in the right
part of (\ref{CEH}) only the first summand in the right part is
non-zero. As a result:
\begin{eqnarray}
H^c_{eff}=\sum \limits_{i=1}^L\omega_n\xi^+_n\xi_n,\label{CEH1}
\end{eqnarray}
and for any \(n=1,2,3...\) \(\omega_n\) is a real positive number.

Operators  \(\xi^+_n,\;\xi_n\) are creation and annihilation
operators of some quasi-particles. As the formula (\ref{CEH1})
indicates, at zero temperature all occupation numbers of these
quasi-particles are equal to zero:  \(n^\xi_l=0,\;l=1,2...\).

Now, we are interested in the average:
\begin{eqnarray}
\langle a^+_n a_n\rangle,\;n=1,2...
\end{eqnarray}
for the ground state of our effective Hamiltonian. Evidently, the
condition \(n^\xi_l=0,\;l=1,2...\) implies \(\forall n=1,2...\)
\begin{eqnarray}
\langle a^+_n a_n\rangle=\sum \limits_{m=1}^{\infty}|{V'}_{nm}|^2,
\end{eqnarray}
where matrix \(V'_{nm}\) is defined according to following equation:
\begin{eqnarray}
a^+_n=\sum \limits_{m=1}^L {U'}_{nm}\xi^+_m+\sum \limits_{m=1}^L
{V'}_{nm}\xi_m.
\end{eqnarray}
Clearly for the interaction of general form \(\hat{U}_2\) between
protein molecules \(\sum \limits_{m,n=1}^L |{V'}_{nm}|^2>0\). This means
that \(\langle a^+_n a_n\rangle>0\) at least for one \(n=1,2...\).

So, if the cell is live, then the number of protein molecules in
the unfolded state is non-zero.

As a conclusion of this section we note that according to above
reasoning, the number of protein molecules in the unfolded state
approaches the finite value, if the cell volume tends to infinity.
The last should mean that the number of the unfolded protein
molecules in the real cell is negligible. The possible solution of
this difficulty is the fact that the protoplasm has a
quasi-crystalline structure only locally within the range of a small
volume we call a domain consisting of several physioatoms.
Therefore, we can supply the stated analysis only within the range
of one domain. Since the domain number in the cell is proportional
to its volume, then the number of protein molecules in the unfolded
state is also proportional to its volume and has anon-zero value
(per unit volume) in the thermodynamic limit.

\section{Discussion of Relation between Mean Field Model and Superfluid
Bose Gas Model on Protein Configuration Space} In the previous
section we quoted the nonlinear Schrodinger equation describing our
system in a mean field approximation:
\begin{eqnarray}
 i\frac{\partial}{\partial
t}\xi=\hat{V}^1\xi+N\hat{V}^2(\xi)\xi,\nonumber\\
\xi \in L_2(X,d\mu),\label{efield}
\end{eqnarray}
where  \(\xi \in L_2(X,d\mu)\), and \(\hat{V}^1,\;\hat{V}^2(\xi)\)
are operators in \(L_2(X,d\mu)\).

The question arises, to which degree the results obtained within the
framework of the mean field method (superfluid model) correlate with
results obtained within the model of the superfluid Bose gas on the
protein configuration space. Now, we'll show that these two models
give the same result for spectrum of single-particle excitations.

Suppose \(\hat{U}_2\) used to construct \(\hat{V}^2(\xi)\) fulfills
the condition which defines, as was said, that all the matrix
elements of \(\hat{U}_2\) such that for them three of
four indexes equal to zero and fourth is non-zero equal to zero. Here we discussed
matrix elements regarding the basis of eigenfunctions of operator
\(\hat{V}^1\).

Since the interaction constant \(\lambda\) is small and almost all
protein molecules are in the normal unfolded state, then \(\xi\) is
given by:
\begin{eqnarray}
\xi=\varphi_0+\eta, \label{efield}
\end{eqnarray}
where \(\eta\) is orthogonal to \(\varphi_0\),
\(\varphi_0,\varphi_1,\varphi_2...\) are an orthonormal basis of
eigenfunctions of operator \(\hat{V}^1\), and \(\varphi_0\)
corresponds to the minimum eigenvalue of \(\hat{V}^1\). The value
\(\eta\) is a first-order value by the coupling constant
\(\lambda\). Let  \(\eta_n\) be Fourier coefficients of vector
\(\eta\) with regard to basis \(\varphi_n\)
\begin{eqnarray}
\eta=\sum \limits_{n=1}^\infty \eta_n\varphi_n.
\end{eqnarray}

If in (\ref{efield}) we keep only terms linear on \(\eta\), then
complex conjugate to (\ref{efield}) has the following form:
\begin{eqnarray}
-i\frac{\partial}{\partial t}\eta_n^{\star}=E_n\eta_n^{\star}+\sum
\limits_{m=1}^{\infty} B_{mn}\eta_m^{\star}+2\sum
\limits_{m=1}^{\infty} A_{mn}^{\star}\eta_m,\;n=1,2,...\label{Bog1}
\end{eqnarray}
A similar derivation can be made for the equation for
\(\frac{\partial}{\partial t}\eta_n\). On the other hand in superfluid model,
after suitable canonical transformation,
\(a^+_n,a_n\), satisfy the Heisenberg equation:
\begin{eqnarray}
\dot{a}^+_n={i[H_{eff},a^+_n]},\;\dot{a}_n={i[H_{eff},a_n]}.
\end{eqnarray}
If we write down these equations in close detail, then we receive:
\begin{eqnarray}
-i\frac{\partial}{\partial t}a_n^{+}=E_na_n^{+}+\sum
\limits_{m=1}^{\infty} B_{mn}a_m^{+}+2\sum
\limits_{m=1}^{\infty} A_{mn}^{*}a_m,\;n=1,2,...\label{Bog2}
\end{eqnarray}
and an equation obtained from the previous one by the Hermitean
conjugation.

But equations (\ref{Bog1}) and (\ref{Bog2}) have the same form,
therefore, if variables \(\{a^+_n,a_n\}\) and
\(\{\eta^+_n,\eta_n\}\) are subjected to linear transformation of
the same form, then equations (\ref{Bog1}) and (\ref{Bog2}) for
transformed variables coincide. Particularly, if the linear
transformation transfers \(\{a^+_n,a_n\}\) into
\(\{\xi^+_n,\xi_n\}\), then \(\eta_n\) transformed under the same
transformation are changed according to the following law:
\begin{eqnarray}
\eta_n=\rm const \mit e^{-i\omega_n t}.
\end{eqnarray}
So \(\eta\) is given by
\begin{eqnarray}
\eta=\sum \limits_{n=1}^{\infty}f_ne^{-i\omega_n t}.
\end{eqnarray}
This formula shows that Bogolyubov's frequencies also describe a
spectrum of single-particle excitations in the mean field model.

\section{Comparison of Van der Waals Protoplasm Model and Superfluid Bose Gas
Model on Configuration Space of Protein Molecule}

The present work considers two models of Ling's cell protoplasm
microstructure.

In the first model called Van der Waals model, we supposed that
nontrivial first integrals of the system are so that their fixing by
predefined values characterizing the resting state of a cell leads
to the fact that protein molecules are situated in points of a
lattice in the unfolded conformation. In addition, if the rapid
descending of Van der Waals interaction between protein molecules
with distance increasing is taken into account, the thermodynamic
features of the living protoplasm can be calculated just as for the
ideal gas of proteins. The thermodynamic features of the dead
protoplasm can be calculated using a well-known Van der Waals
interpolation formula for the free energy of a system. With this
model we've obtained an expression for the quantity of heat the cell
released while dying, and for a number of potassium ions releasing
from the cell during this process.

In the second model we emphasize the analysis of internal protein
molecules structure. As we have already mentioned, the protein
molecules are supposed to be situated in points of a lattice, but
have nontrivial internal degrees of freedom; and we study the
structure of the ground state of this molecular system. The present
work shows that within the range of weak interaction between protein
molecules, the ground state of the interacting proteins system
should be outlined as a ground state of the Hamiltonian of the Bose
gas with a weak interaction on the protein molecule configuration
space. To analyze this Hamiltonian we used standard methods of the
superfluid theory. Taking into account that the interaction between
protein molecules is weak and the bulk of protein molecules of the
system is in the ground folded state, our Hamiltonian can be
replaced with an effective quadratic one, but it has different forms
depending on if the cell is live or dead. In addition, it clears up
which nontrivial motion integrals in the involution should be
included to the generalized Gibbs distribution describing the living
cell. The use of these effective quadratic Hamiltonians reveals that
in the dead cell all the protein molecules significant for model
properties are folded, while in the resting living cell the number
of unfolded protein molecules is non-zero.

The question arises: what is the relation between considered
protoplasm models? Particularly, how do conservation laws postulated
for them link? We think, the considered models complement each other
and connection of the named conservation laws comes down to the
following. If we fix the motion integrals for the second model by
some predefined values, then protein molecules are in the unfolded
state, and we can suppose that the nature of interaction between
different protein molecules provides the formation of a lattice
being an energetically favorable structure. In other words, the
conservation laws for the first model appear to be consequences of
the conservation laws for the second model. If we conversely
consider the Van der Waals protoplasm model as the main one and
assign conservation laws to it, according to which the protein
molecules in the living state are situated in points of a lattice,
then it is reasonable to suggest that such a space configuration of
protein molecules would be favorable for the unfolded (not folded)
state of protein molecules for a number of reasons. In other words,
the conservation laws for the second model are consequences of the
conservation laws for the first model. This situation is certainly
very inexact and hypothetical.

The questions, on which principles should new models be
constructed(except ones considered here) and how to develop the
kinetic theory of the living cell, are the subject of further
investigations.

\section{Conclusion}
In the present work we have constructed and investigated properties
of two complementary models of protoplasm --- physical basis of
life. The work was realized basing on generalized thermodynamics we
proposed (Prokhorenko and Matveev, 2011). Within the framework of
stated assumptions we explained a number of properties and phenomena
observed in living cells, formalized (in the extended sense) by the
physical theory of the living cell by Ling (2001). However, the
results we had formerly obtained, were inequalities and denoted only
processes direction (in excitation and death of a cell heat
releases, volume reduces and so on), there were no certain numerical
evaluations obtained. To obtain them we need to specify the
properties of intermolecular interactions which escape from the
point of view of available analyze methods. That's why, we should
construct different protoplasm models emphasizing one system
parameter after another.

In the present work we constructed and investigated two models of
protoplasm: one of them is naturally called Van der Waals model, and
other is the model of superfluid Bose gas on the protein molecule
configuration space. Our aim in this work was not to construct a
model giving the most exact agreement with experimental data but to
show that the constructing of such models is reasonable and
possible. Qualitative agreement of the obtained results with
experimental data gives an evidence of vitality of the thermodynamic
theory of the living cell we proposed (Prokhorenko and Matveev,
2011).

Our theory can face the misapprehension from devotees of the
non-equilibrium thermodynamics, as it is based on the equilibrium
statistical physics and thermodynamics. Their main argument is
evident: in the equilibrium state the maximum entropy is reached,
therefore, the cell cannot perform the biological work. However, the
other fact is also evident: the non-equilibrium thermodynamics still
haven't given the quantitative description even for elementary
phenomena, for example, electric resting potential. In addition, if
we accept that the cell is living under the laws of non-equilibrium
thermodynamics, we must also recognize that 200 years experience in
successful modeling properties of the cell by non-living systems is
untenable in principle. It would ruin all our knowledge about the
living cell without giving anything in return. As for our
generalized thermodynamics, the essential clarification is required:
though the states which it operates are static by time, they are not
equilibrium in the sense that their entropy is not the maximum one
of all the states having the same energy (this is an essential
feature of the state of the living material). The existence of such
states even for the most realistic statistical mechanics systems is
proven by one of us (Prokhorenko, 2009).

We are very grateful to P. Agutter, A.V. Koshelkin, Yu. E. Lozovik,
A.V. Zayakin, E.N. Telshevskiy and N.V. Puzyrnikova for valuable critical
comments on this article and very useful discussions.

\section{Appendix 1. Transition of Cell from Living to Dead State in
Weak External Alternating Field} Here and in our previous work
(Prokhorenko and Matveev, 2011) we considered two extreme states of
a cell: non-equilibrium state of rest and equilibrium state
corresponding to death. Now, let's ask a question, can we describe
death of the protoplasm within the framework of our generalized
thermodynamics, i.e. to construct an example of a process
transferring a cell from the living state of rest to the state
corresponding to death. If we could not construct an example of such
a process, all our theory would appear to be doubtful. In addition, up
till now we considered only equilibrium statistical mechanics of the
living cell, and the constructing of an example of the
transformation process is the first step toward the construction of
the non-equilibrium statistical mechanics of the living cell.

Let's start the analysis of the cell death mechanism from a certain
problem of the non-equilibrium statistical mechanics. Let's consider
the Hamilton system having \(n\) degrees of freedom which is
described by the Hamiltonian \(H(p, q)\) in the field of external
forces \(\varepsilon f(t),\;\varepsilon \in \mathbb{R}\) such that
the complete Hamiltonian of the system is as follows:
\begin{eqnarray}
\Gamma=H+\varepsilon f(t) P,
\end{eqnarray}
where \(P(p,q)\) is a function of canonically conjugated momenta and
coordinates of a system. We suppose that applied force \(f(t)\) is a
real function of time of the following form:
\begin{eqnarray}
f(t)=\sum \limits_{\nu}a_\nu \cos(\nu t+\varphi_\nu),
\end{eqnarray}
where phases \(\varphi_\nu\) are independent random values uniformly
distributed by a circle. We suppose the frequency spectrum of the
applied force is essentially continuous, so that sums of the
following form
\begin{eqnarray}
\sum \limits_{\nu}F(\nu)a_\nu^2
\end{eqnarray}
with continuous \(F(\nu)\) in the limit can be replaced by integrals
\begin{eqnarray}
\int \limits_{0}^{+\infty}F(\nu)I(\nu)d\nu.
\end{eqnarray}
This problem was considered by N.N. Bogoliubov (1945) and here we
briefly quote the results obtained by them (relevant only to
classical mechanics).

Let's denote \(D_t\) is a probability density of coordinates and
momenta in a time \(t\) provided that phases \(\varphi_\nu\) have
some defined values. The probability density for distribution \(p\)
and \(q\) in the common sense, i.e. when phase values
\(\varphi_\nu\) are inessential, can be found from \(D_t\) by
averaging over all phases:
\begin{eqnarray}
\rho_t=\overline{D_t}.
\end{eqnarray}
In the initial moment \(t = 0\) we suppose that distribution of
coordinates and momenta does not depend on phases, in other terms
\begin{eqnarray}
D_0=\rho_0.
\end{eqnarray}

Temporal evolution of \(D_t\) should be passed in accordance with a
well-known Liouville's equation:
\begin{eqnarray}
\frac{\partial D_t}{\partial t}=(\Gamma,D_t)+\varepsilon
f(t)(P,D_t),
\end{eqnarray}

where \((A,B)\) is a Poisson bracket defined by the following
formula:
\begin{eqnarray}
(A,B)=\sum \limits_{i=1}^n(\frac{\partial A}{\partial
q_i}\frac{\partial B}{\partial p_i}-\frac{\partial B}{\partial
p_i}\frac{\partial A}{\partial q_i}).
\end{eqnarray}

Let's introduce more one-parameter group of operators \(T_t\; t\in
\mathbb{R}\) acting on dynamical variables according to the formula:
\begin{eqnarray}
T_tF(p,q)=F(p_t,q_t),
\end{eqnarray}
where \(p_t,q_t\) is a solution of canonical Hamilton equations
\begin{eqnarray}
\frac{d p_t}{dt}=-\frac{\partial H(p,q)}{\partial q}, \nonumber\\
\frac{d q}{d t}=\frac{\partial H(p,q)}{\partial q},
\end{eqnarray}
having initial data
\begin{eqnarray}
p_0=p,\; q_0=q.
\end{eqnarray}

Using these assumptions and designations (Bogoliubov, 1945), in the
limit of small \(\varepsilon\) the following equation for \(\rho_t\)
was derived:
\begin{eqnarray}
\frac{\partial \rho_t}{\partial t}=[H,\rho_t]+\varepsilon^2\int
\limits_{0}^{t} \Delta (t-\tau)(P,(T_{\tau-t}(P,\rho_\tau))d\tau,
\label{12}
\end{eqnarray}
where
\begin{eqnarray}
\Delta(\tau)=\frac{1}{2} \int \limits_{0}^{+\infty} I(\nu)
\cos(\nu\tau) d\tau.
\end{eqnarray}
Further Bogoliubov (1945) considered the case when the system
exposed to external force is a harmonic oscillator of
\(n\)-dimensions having incommensurate frequencies
\(\vec{\omega}=(\omega_1,...,\omega_n)\). Let's denote through  a
\(\sigma_t=\overline{\rho_t}\) function of action variables obtained
from \(\rho_t\) by averaging by angular variables
\(\theta_1,...,\theta_n\). Function \(P(p,q)\) becomes a function of
action-angle variables and can be expanded to the Fourier series:
\begin{eqnarray}
P=\sum \limits_{\vec{n}}P_{\vec{n}} e^{i\vec{\theta}\cdot \vec{n}}.
\end{eqnarray}
In this case, at the same initial conditions on \(D_t\), in the limit
of small \(\varepsilon\) the evolution of \(\sigma_t\) is described
by the Fokker-Planck equation:
\begin{eqnarray}
\frac{\partial \sigma_t}{\partial t}=\varepsilon^2 \sum
\limits_{k=1}^{n} \sum \limits_{s=1}^n \frac{\partial}{\partial
I_s}A_{sk}(I)\frac{\partial \sigma_t}{\partial I_k},
\end{eqnarray}
where
\begin{eqnarray}
A_{sk}(I)=\frac{\pi}{4} \sum \limits_{\vec{n}} n_k n_s I(\vec{n}
\cdot \vec{\omega})|P_{\vec{n}}(I)|^2.
\end{eqnarray}
This equation, particularly, describes diffusion of distributions
which at the initial moment can be distributions concentrated in one
point, i.e.
\begin{eqnarray}
\sigma_0(t)=\frac{1}{(2\pi)^n}\prod \limits_{k=1}^n
\delta(I_k-I_k^0).
\end{eqnarray}
But the last distribution is a generalized microcanonical
distribution which we used to describe the states of rest of the
Ling's cell. Therefore, it seems to be natural if the method proposed
by  Bogoliubov (1945) can be used to construct an example of a cell
transformation process from the living to dead (equilibrium) state.
Now, we shall show how it can be done using the results of Appendix 1
of the work by (Prokhorenko and Matveev,
2011).

So, assume there is a Hamilton system \(M\) with \(n+k\) degrees of
freedom where \(k\ll n\), which is described by the Hamiltonian \(H\) and where
\(k\) independent first integrals in the involution are defined.
There was shown in the Appendix 1 of the above-mentioned work that
some covering Hamiltonian system \(M'\) of system \(M\) can be
represented as a direct product of \(M'=M'_1\times M'_2\), so that
the canonically conjugated coordinates on \(M'_2\) are first
integrals  \(K'_1,...,K'_k\) (more properly, their lifting to
\(M'\)) playing a role of momenta and "angle" variables
\(\varphi_1,...,\varphi_k\), passing all the real axis and playing a
role of coordinates.

Let \(\pi\) be a canonical projection of \(M'\) on \(M\). The
function \(f\) on \(M'\) we (Prokhorenko and Matveev, 2011) called a
periodic one if \(f(x)=h(\pi(x))\) for some function \(h\) defined
on \(M\). It has been suggested (Prokhorenko and Matveev, 2011) to
represent the mixed state of the \(M'\) system using some positive
periodic function \(\rho\) on \(M'\). Put by definition \(R_L:=\{x
\in M'||\varphi_1|<L,...,|\varphi_k|<L\}\). If \(A\) is a periodic
function on \(M'\), then its average over the state corresponding to
distribution \(\rho\) was suggested (Prokhorenko and Matveev, 2011)
to calculate using the following formula:
\begin{eqnarray}
\langle A\rangle=\lim_{L\rightarrow \infty} \frac{\int \limits_{R_L}
\rho(
K_1,...,K_k,\varphi_1,...,\varphi_k)A(K_1,...,K_k,\varphi_1,...,\varphi_k)dK_1...d\varphi_k}{\int
\limits_{R_L} \rho(
K_1,...,K_k,\varphi_1,...,\varphi_k)dK_1...d\varphi_k}.\label{11}
\end{eqnarray}
In this work we have shown that the limit (\ref{11}) always exists.
We shall call the distribution function
\(\rho(K_1,...,K_k,\varphi_1,...,\varphi_k)\) as a normalized one if
\begin{eqnarray}
\lim_{R_L\rightarrow \infty}\frac{1}{L^k}\int \limits_{R_L}\rho(
K_1,...,K_k,\varphi_1,...,\varphi_k) dK_1...d\varphi_k=1.
\end{eqnarray}
The entropy of a state corresponding to the normalized distribution
\(\rho(K_1,...,K_k,\varphi_1,...,\varphi_k)\) is defined by the
following formula
\begin{eqnarray}
S=-\lim_{R_L\rightarrow \infty}\frac{1}{L^k}  \int
\limits_{R_L}\rho( K_1,...,K_k,\varphi_1,...,\varphi_k) \ln \rho(
K_1,...,K_k,\varphi_1,...,\varphi_k) dK_1...d\varphi_k. \label{14}
\end{eqnarray}
The last limit exists while the proof of its existence is the same
as for the limit (\ref{11}).

The Hamiltonian of the covering system \(M'\) we also (without a
risk to make an error) denote as \(H\). If our Hamilton system is a
Ling's cell, then  \(n\gg k\). In this case, as shown (Prokhorenko
and Matveev, 2011), the dynamics of the system \(M'_2\) can be
considered separately. This dynamics is a Hamiltonian one and is
defined by Hamiltonian \(F(x| T)\) where \(F(x| T)\) is the free
energy of the system \(M'_1\) at temperature \(T\) provided that the
\(M'_2\) system is situated in point \(x\):
\begin{eqnarray}
F(x|T)=-T \ln \int d\Gamma_y^1 e^{-\frac{H(y,x)}{T}},
\end{eqnarray}
\(y \in M'_1\), \(x \in M'_2\), \(d\Gamma^1_y\) is an element of
phase volume on \(M'_1\).

But  \(K_1,...,K_k\) are the motion integrals, therefore \(F(x|T)\),
does not depend on angular variables \(\varphi_1,...,\varphi_k\),
but depends only on \(K_1,...,K_k\). Put by definition
\begin{eqnarray}
F'(K_1,...,K_k|T):=F(x|T).
\end{eqnarray}
Further, we are interested in the case when \(F'(K_1,...,K_k)\)
reaches its minimum in the whole area (of non-zero volume)
\(\mathcal{O}\). In the work by Prokhorenko and Matveev (2011) the corresponding thesis is called the
equivalence principle and there was shown that only under condition
of \(F'(K_1,...,K_k)\) constancy in the whole area \(\mathcal{O}\)
of the non-zero volume our generalized thermodynamics gives new
results in comparison with the common one and can provide the
thermodynamic descriptions of resting state of the Ling's cell.

So, \(F(x|T)\) is a constant on the direct product
\(\mathcal{O}\times\mathbb{R}^k\) and, therefore, it defines a
trivial dynamics in this area, i.e. if \((p_0,q_0)\in
\mathcal{O}\times\mathbb{R}^k\), then \(\forall t \in \mathbb{R}\)
\((p_t,q_t)=(p_0,q_0)\).

Using the last circumstance the equation (\ref{12}) can be
simplified as follows
\begin{eqnarray}
\frac{\partial \rho_t}{\partial t}=\varepsilon^2\int \limits_{0}^{t}
\Delta (t-\tau)(P,(P,\rho_\tau))d\tau.
\end{eqnarray}
However, in the limit of small \(\varepsilon\) \(\rho_\tau (z)\)
varies with time slowly and is almost constant wherever
\(\Delta(t-\tau)\) noticeably differs from zero. Therefore, in the
integral from the right part of the last equation we can replace
\(\rho_\tau\) by its value at \(\tau=t\). As
\begin{eqnarray}
\int \limits_0^{+\infty}\Delta(t)dt=\frac{\pi I(0)}{2}
\end{eqnarray}
we can find
\begin{eqnarray}
\frac{\partial\rho_t}{\partial t}=\frac{\varepsilon^2\pi
I(0)}{2}(P,(P,\rho_t)). \label{13}
\end{eqnarray}
This equation of Fokker-Planck type is correct in the area
\(\mathcal{O}\times\mathbb{R}^k\).

As for behavior of function \(\rho_t\) outside the area
\(\mathcal{O}\times\mathbb{R}^k\), the physical considerations make
clear that the probability to find the system \(M'\) outside the
domain \(\mathcal{O}\times\mathbb{R}^k\) is negligible. Therefore, it
is sufficient to investigate the behavior of \(\rho_t\) inside the
domain \(\mathcal{O}\times\mathbb{R}^k\) by applying the appropriate
boundary conditions for \(\rho_t\) on the boundary of
\(\mathcal{O}\times\mathbb{R}^k\) which provide the self-adjointness
of equation (\ref{13}) and keeping the probability:
\begin{eqnarray}
\int \limits_{\mathcal{O}\times\mathbb{R}^k}\rho_t(x)d\Gamma^2_x=1,
\end{eqnarray}
where \(d\Gamma^2_x\) is an element of phase volume on \(M'_2\).
However, we won't make an ascertaining of the form of these boundary
conditions but just make an assumption that the domain
\(\mathcal{O}\) is a compact manifold (without boundary).

The self-adjointness condition of operator from the right part of
the Fokker-Plank equation is a consequence of equality of forward
and backward probabilities of arbitrary transition. This equality of
forward and backward probabilities of transitions is a consequence
of complete Hamiltonian invariance with respect to the time sign
conversion operation (See the principle of kinetic coefficients
symmetry by L. Onsager).

Now, following Bogolyubov (1945), we can show that the entropy of a
state defined by relation (\ref{14}) is increased monotonically in
course of time due to equation (\ref{13}). The entropy (\ref{14})
reaches its maximum at the constant distribution function, i.e. at
\(t=+\infty\)
\begin{eqnarray}
\rho_{+\infty}(K_1,...,\varphi_k)=\rm const \mit.
\end{eqnarray}

But if the distribution function
\(\rho_{+\infty}(K_1,...,\varphi_k)\) is constant, then it
corresponds to the Gibbs equilibrium microcanonical distribution.

So, an example of the Ling's cell transformation from the living
(resting) to the dead state is its evolution in the weak external
field which is a sum of harmonic oscillations with continuous
frequency spectrum and random independent phases uniformly
distributed along the circle; at that, for spectral density of
intensity of this field \(I(\nu)\), \(I(0)\neq 0\) is required to be
fulfilled.

The fact that the Fokker-Planck equation includes the spectral
density of intensity only through \(I(0)\) is very significant and
can be verified experimentally. This conclusion is in a good
agreement with observed destructive influence of infrasound,
magnetic storms and any white noises to living organisms.

\section{Appendix 2. On ATP Structuring Role as Part of (ATP)m-(Protein)n-(H2O)p-(\({\rm K \mit}^+\))q Complex}
In this Appendix we discuss the theoretical relation between the
capability of an ATP molecule to define the structure of a
physioatom in the resting state and non-ergodicity of the Ling's
cell. The purpose of this Appendix is to represent a demonstrative
physical view of the structuring ATP influence on the physioatom. We
do not give a mathematically consistent characteristic of
ATP-protein-water-\({\rm K \mit}^+\) interaction here.

We are interested in the following issues. How much is a number of
protein molecules in the water-protein complex managed by one ATP
molecule? Why the physical disturbance generated by the ATP
propagates without dissipation to the large number of proteins and
how does this disturbance influence on their conformation?

The fact that, according to our main assumption, the Ling's cell
represents a Hamiltonian system having a vast number of first integrals
in the involution makes us think, can we use the theory of
completely integrable (in the sense of Liouville) systems to
describe such a cell? The Korteweg---de Vries equation (Arnold,
2003) which can be an example of a completely integrable system
\begin{eqnarray}
u_t=6uu_x-u_{xxx}
\end{eqnarray}
was originated in the shallow water theory (in narrow ship
channels). This equation is remarkable by allowing the solutions in
a form of solitary waves (solitons) propagating without dissipation.
The only parameter characterizing the soliton is its velocity. The
Korteweg---de Vries equation also allows for multi-solitonic
solutions which break into separate solitons propagating with
different velocities at \(t\rightarrow\pm \infty\). A significant
property of multi-solitonic solutions is the fact that in case of
solitons collision their velocities do not change.

We consider that the distribution of the ATP molecule physical
influence on surrounding protein molecules has something similar to
the propagating of solitons because there are many commutative first
integrals for the Ling's cell, therefore, this case should be
somewhat similar to the case arising in the theory of completely
integrable systems.

So that one ATP molecule could effectively manage the surrounding
complex of water and proteins, the disturbance transferred by
solitons should not dissipate. For analysis simplicity let's suppose
that the distribution of the physical impulse from the ATP is
described by a one-dimensional Korteweg---de Vries equation. Then
every soliton is unambiguously defined by the only parameter, its
velocity. Therefore, to prevent loss of information from the ATP
molecule, solitons' velocities shall not change after their
collision. But the last property is fulfilled for multi-solitonic
solutions of the Korteweg---de Vries equation, as we've mentioned
above.

The theory of completely integrable systems is a rather developed
one (Bullaf and Caudry, 1999). A lot of integrable equations were
constructed on the line, for example: nonlinear Schrodinger
equation, sine-Gordon equation, Toda chain and so on. The main
method of integrating such equations is a method of inverse
scattering problem (Bullaf and Caudry, 1999). The property of
isolated solutions to pass through each other without velocity
changes can be common for all of them and is explained by the
presence of the complete set of independent commutative first
integrals.

Let's take, for instance, the finite Toda chain consisting of \(N\)
particles in the line (Mozer, 1975). The state of this system is
fully described by defining \(N\) particle coordinates
\(\{x_i|i=1,...,N\}\) and \(N\) momenta \(\{p_i|i=1,...,N\}\). By
definition the Hamiltonian of this system is:
\begin{eqnarray}
H=\sum \limits_{i=1}^N \frac{p_i^2}{2}+\sum
\limits_{i=1}^{N-1}e^{x_{i+1}-x_i}.
\end{eqnarray}
As shown by Mozer (1975), at \(t\rightarrow\pm\infty\) the distances
between different particles tend to infinity. This system is
completely integrable (Mozer, 1975) and there is a set of \(N\)
independent first integrals in the involution for this system. If
distances between particles are so large so their interaction can be
neglected, then those integrals are just elementary symmetrical
polynomials of momenta (velocities) (Mozer, 1975).
\begin{eqnarray}
I_1=p_1+p_2+...+p_N,\nonumber\\
I_2=p_1p_2+...+p_1p_N+..+p_{N-1}p_N\nonumber\\
........................................................\nonumber\\
I_3=p_1p_2...p_N
\end{eqnarray}
As before and after particles' collisions the values of integrals
\(I_1,...,I_N\) coincide, to define the velocities after collision
we have a system of \(N\) algebraic equations which implies that
velocities of particles after collision are the roots of algebraic
equation
\begin{eqnarray}
(v-v_1)...(v-v_N)=0,
\end{eqnarray}
where \(v_1,...,v_N\) are velocities of particles before collision
and \(v\) is an unknown variable. So, velocities of particles after
collision coincide with velocities of particles before collision
with an accuracy of transmutation.

Yet notice that for integrable systems with a number of freedom
degrees \(N\rightarrow \infty\) the equivalence principle is
fulfilled. Suppose \(I_1,...,I_N\) are action variables, and
\(\varphi_1,...,\varphi_N\) are angle variables conjugated to them.
We can take \(I_2,...,I_N\) as independent first integrals in the
involution, which are included in the generalized microcanonical
distribution. We have:
\begin{eqnarray}
S(E,I_2,...,I_N)=\ln \int dI_1 d\varphi_1 \prod \limits_{i=2}^N d
\varphi_i \delta(H(I_1,...,I_N)-E).
\end{eqnarray}
Integrating this formula by \(dI_1\) results in:
\begin{eqnarray}
S(E,I_2,...,I_N)=\ln \frac{1}{|\frac{\partial
H(I_1,...,I_N)}{\partial I_1}|}+N \ln 2\pi.
\end{eqnarray}
But \(\omega_1(I_1,...,I_N)\) is a frequency corresponding to
\(\varphi_1\); and it is reasonable to restrict our choice to
systems for which \(\omega_1\) is an asymptotically constant at
\(N\rightarrow \infty\). So:
\begin{eqnarray}
S(E,I_2,...,I_N)=-\ln \omega_1(I_1,...,I_N)+N \ln 2\pi
\end{eqnarray}
But since \(\omega_1(I_1,...,I_N)\) is an asymptotically constant,
in the limit of \(N \rightarrow \infty\) can be neglected, and the
equivalence principle is fulfilled for integrals \(I_2,...,I_N\) in
the limit of \(N \rightarrow \infty\).

\end{document}